\documentclass[letterpaper,12pt]{article}
\usepackage{color,amssymb,enumerate,float}

\usepackage{ifpdf}
\ifpdf
	\usepackage[pdftex]{graphicx}
	\usepackage[pdftex,unicode,implicit]{hyperref}

	\hypersetup{
  	pdftitle     = {Wormholes in Horava gravity with cosmological constant},
  	pdfkeywords  = {},
  	pdfauthor    = {J. Bellor\'{\i}n, A. Restuccia and A. Sotomayor},
  	pdfcreator   = {pdf\LaTeXe\ with package \flqq hyperref\frqq},
  	pdfproducer  = {pdf\LaTeXe\ with package \flqq hyperref\frqq},
  	pdfpagemode  = UseNone,  
  	pdffitwindow = true,  
  	unicode      = true,
  	plainpages   = true,
  	colorlinks   = true,  
  	citecolor    = blue,  
  	urlcolor     = blue,
  	linkcolor    = blue
	}

	\newcommand{\hepth}[1]{\hspace*{-0.5ex}
		{\href{http://www.arXiv.org/abs/hep-th/#1}{arXiv:hep-th/#1}}}

	\newcommand{\grqc}[1]{\hspace*{-0.5ex}
		{\href{http://www.arXiv.org/abs/gr-qc/#1}{arXiv:gr-qc/#1}}}

	\newcommand{\arXiv}[1]{\hspace*{-0.5ex}
		{\href{http://www.arXiv.org/abs/#1}{arXiv:#1}}}

\else

  \usepackage[dvips]{graphicx}
  \usepackage[unicode,implicit]{hyperref}
  \newcommand{\hepth}[1]{{arXiv:hep-th/#1}}
  \newcommand{\grqc}[1]{{arXiv:gr-qc/#1}}
  
  \newcommand{\arXiv}[1]{{arXiv:#1}}

\fi

\makeatletter
\@addtoreset{equation}{section}
\makeatother

\begin{document}

\begin{center}

\vspace*{2cm}
{\bf \LARGE Wormholes in Ho\v{r}ava gravity with cosmological constant} 
\vspace*{2cm}

{\sl\large Jorge Bellor\'{\i}n,}$^{a,}$\footnote{\tt jorgebellorin@usb.ve}
{\sl\large Alvaro Restuccia}$^{a,b,}$\footnote{\tt arestu@usb.ve}
{\sl\large and Adri\'an Sotomayor}$^{c,}$\footnote{\tt asotomayor@uantof.cl}
\vspace{3ex}

$^a${\it Department of Physics, Universidad Sim\'on Bol\'{\i}var, Valle de Sartenejas,\\ 
1080-A Caracas, Venezuela.} \\[1ex]
$^b${\it Department of Physics}, $^c${\it Department of Mathematics, Universidad de Antofagasta, 1240000 Antofagasta, Chile.}

\vspace*{2cm}
{\bf Abstract}
\begin{quotation}{\small By combining analytical and numerical methods we find that the solutions of the complete Ho\v{r}ava theory with negative cosmological constant that satisfy the conditions of staticity, spherical symmetry and vanishing of the shift function are two kinds of geometry: (i) a wormhole-like solution with two sides joined by a throat and (ii) a single side with a naked singularity at the origin. We study the second-order effective action. We consider the case when the coupling constant of the $(\partial\ln{N})^2$ term, which is the unique deviation from general relativity in the effective action, is small. At one side the wormhole acquires a kind of deformed AdS asymptotia and at the other side there is an asymptotic essential singularity. The deformation of AdS essentially means that the lapse function $N$ diverges asymptotically a bit faster than AdS. This can also be interpreted as an anisotropic Lifshitz scaling that the solutions acquire asymptotically.
}\end{quotation}

\end{center}

\thispagestyle{empty}

\newpage
\section{Introduction}
The nonrenormalizability of general relativity (GR) in the perturbative scheme as well as the issues of dark matter and dark energy open the doors for the possibility of studying completions of GR. In this sense the Ho\v{r}ava proposal \cite{Horava:2009uw} of having an ultraviolet completion of GR with a preferred foliation of space-time has been studied widely. Among the several versions that have been proposed, the complete nonprojectable version \cite{Horava:2009uw,Blas:2009qj} exhibits a good behavior in what concerns to the structure of the field equations and constraints \cite{Blas:2009qj,Donnelly:2011df,Bellorin:2011ff,Bellorin:2012di,Bellorin:2013zbp}. For this version most of the analysis has been done outside the conformal point of the kinetic term, which holds at the value $\lambda = 1/3$ of its corresponding coupling constant. Out from the conformal point, $\lambda \neq 1/3$, the theory propagates one extra mode in addition to the two tensorial modes of GR\footnote{An exception (at the level of classical actions) is the model with only a $R$-term in the potential, which is equivalent to GR for all values of $\lambda$ and for asymptotically flat geometries. This was shown in \cite{Bellorin:2010je} and corroborated in \cite{Loll:2014xja,Das:2011tx}.}. The decoupling of this extra mode at low energies faces with the so-called strong coupling problem \cite{Charmousis:2009tc,Papazoglou:2009fj,Blas:2009ck,Kimpton:2010xi}. However, at the conformal point of the kinetic term, $\lambda = 1/3$, the theory propagates exactly the same modes of GR as a consequence of additional second-class constraints that arise at this point \cite{Bellorin:2013zbp} (see also \cite{Park:2009hg}). This is a outstanding feature, since there is no need of any decoupling mechanism and there are no discontinuity issues at low energies. This suggests that theory can be smoothly connected to GR \cite{Bellorin:2010je,Bellorin:2013zbp}. Because of this we consider the $\lambda = 1/3$ case of the nonprojectable theory as a interesting model among the various proposals known for quantum gravity.

Mainly because of their potential astrophysical applications, in the modifications of GR it is mandatory to identify solutions representing isolated sources. A suitable sector to start with is the set of spherically symmetric and static configurations. Obviously, the most important information for these configurations comes from the large-distance effective theory. Thus, our main interest in this paper is to study static spherically symmetric solutions of the effective theory of the complete nonprojectable Ho\v{r}ava theory. Since these configurations are static (and since we switch off the shift function), they do not depend on the constant $\lambda$, thus the configurations are the same for the $\lambda = 1/3$ and $\lambda \neq 1/3$ cases.

The static spherically symmetric solutions with vanishing shift function of the effective theory without cosmological constant were found in Ref.~\cite{Eling:2006df} (see also \cite{Eling:2003rd}). Actually, the solutions of Ref.~\cite{Eling:2006df}  were found in the context of the Einstein-aether theory \cite{Jacobson:2000xp}. The correspondence comes from the equivalence this theory has with the second-order effective action of the Ho\v{r}ava theory \cite{Blas:2009ck,Jacobson:2010mx,Jacobson:2013xta}. An independent analysis of the same configurations but directly on the Ho\v{r}ava theory was done in Ref.~\cite{Kiritsis:2009vz}, where several features of the solutions were studied. Later on, we considered again the problem of the static spherically symmetric solutions of the complete nonprojectable Ho\v{r}ava theory in Ref.~\cite{Bellorin:2014qca} (with vanishing shift function). The main feature of the solutions (for positive mass) is that they have a wormhole-like geometry with a throat joining the two sides. They have an asymptotically flat side and an essential singularity at the infinite boundary of the other side. In Ref.~\cite{Bellorin:2014qca} we recovered the same solutions of Ref.~\cite{Eling:2006df} but explicitly on a coordinate system valid at the throat. In particular this allows to show directly that the physical radius $r$ has a minimum at the throat.

The results of our previous work \cite{Bellorin:2014qca} encourage us to consider the problem of finding the static spherically symmetric solutions of the Ho\v{r}ava theory \emph{now with a cosmological constant turned on}. Specifically, we shall consider the case of negative cosmological constant, under which we can get better control of the behavior of the field equations. As we have already mentioned, for the analysis we take the lowest-order effective action, which contains the cosmological constant and the terms of second order in derivatives. We comment that in Ref.~\cite{Kiritsis:2009vz} it was also studied the same effective action with negative cosmological constant, with focus on the asymptotic behavior of the solutions. In this paper we go beyond the results of \cite{Kiritsis:2009vz} by performing analysis of the solutions over all the space with the aim of understanding their whole geometry. Our program is the implementation of a procedure parallel to the one of the $\Lambda = 0$ case we previously studied \cite{Bellorin:2014qca}. In particular we pursue a new radial coordinate that allows to identify the geometry of the solutions. We shall arrive at differential equations determining both the new coordinate and the metric components in terms of it. With analytical techniques we shall extract important properties from these equations, such as the presence of a minimum for the physical radius $r$, which indicates that the geometry has a throat. Since in this case the relevant equations are considerably more involved than in the $\Lambda = 0$ case, eventually we shall perform numerical integration on them. This will confirm the analytical results and complete the understanding of the configurations.

It is worth stressing that, because the smaller group of symmetries and the different dynamics, in Ho\v{r}ava theory one does not expect to find spherically symmetric configurations possessing the universality properties that the Schwarzschild solution has in GR. To start with, in (vacuum) Ho\v{r}ava theory spherical symmetry does not imply staticity or asymptotic flatness. Therefore, in our study we impose spherical symmetry and staticity independently. We also explore the asymptotia of the solutions, both analytically and numerically, rather than imposing it. Moreover, once the general ansatz satisfying spherical symmetry and staticity has been put in spherical and adapted-time coordinates, it still has the degree of freedom of the shift function open. Specifically, the time-radial component is left free and it can not be gauge-fixed. Different forms of this component lead in general to inequivalent configurations in the context of the Ho\v{r}ava theory. For the sake of simplicity we switch off this function. By doing so we obtain a definite kind of solutions, but the reader should take into account that there are other inequivalent solutions that are static and spherically symmetric. In particular, black holes have been found under such conditions in the complete nonprojectable Ho\v{r}ava theory \cite{Barausse:2011pu,Blas:2011ni}.\footnote{Since in Refs.~\cite{Barausse:2011pu,Blas:2011ni} the theory is formulated in a covariant way with an additional vector field, the extra degree of freedom arises as the radial component of the vector field rather than in the shift vector.} Universal horizons which arise in solutions like those have been the subject of intense study, see for example \cite{various:universalhorizon} and references therein. Directly in the original Ho\v{r}ava theory without the terms of Ref.~\cite{Blas:2009qj}, solutions with the radial component of the shift function turned on have also been found \cite{Capasso:2009ks}. Other exact solutions of various versions of the Ho\v{r}ava theory can be found for example in Refs.~\cite{Kiritsis:2009rx,Lu:2009em,Kehagias:2009is}. In particular, in Ref.~\cite{BottaCantcheff:2009mp} wormhole configurations were considered in the original Ho\v{r}ava theory without the terms of Ref.~\cite{Blas:2009qj}.

The study we present here can be also of interest for holography. Solutions of Ho\v{r}ava theory has been used as holographic duals of Lifshitz-scaling field configurations; see, for example, Ref. \cite{Griffin:2012qx}. Since we study the theory with negative cosmological constant, the spacetimes we study here could enter into the arena of holographic duals (at least numerically).

As we have mentioned, another important aspect we shall investigate is the asymptotic behavior of the solutions. Under a hypothesis of polynomial divergence, which can be posed even before finding the global structure, we shall determine analytically whether the solutions are asymptotically anti-de Sitter and we shall contrast the analytical result with the numerical solution. We give in advance that the asymptotia we find (at one of the sides of the throat) is not exactly anti-de Sitter, but rather a deformation of it. At the other side there is an essential singularity similar to the one found in the inner side of the $\Lambda = 0$ case \cite{Eling:2006df,Bellorin:2014qca}.


\section{Solutions of the field equations}
\subsection{The field equations and the ansatz}
In Ho\v{r}ava theory a given foliation of spacetime is considered of absolute physical meaning, such that the allowed symmetry transformations can not change the foliation \cite{Horava:2009uw}. The foliation can be defined by the existence of a global ``time" function and for convenience the time coordinate $t$ can be identified with this function. Then the spacetime is foliated by hypersurfaces of $t=\mbox{constant}$ which are spacelike. In this context it is convenient to parameterize the spacetime metric in terms of the Arnowitt-Deser-Misner (ADM) variables, such that
\begin{equation}
 ds^{2} = - (N^2 - N_i N^i) dt^2 + 2 N_i dx^i dt + g_{ij} dx^i dx^j \,.
\end{equation}
We assume that the lapse function $N$ depends both on time and space, hence we deal with the nonprojectable formulation of the theory \cite{Horava:2009uw}. In particular, this assumption is fundamental to find the kind of configurations we are interested in. The most general kind of diffeomorphisms over the spacetime that do not change the given foliation are the foliation-preserving diffeomorphisms, under which the coordinates and ADM variables transform as
\begin{equation}
\begin{array}{l}
\delta t = f(t) \,,
\hspace{2em}
\delta x^i = \zeta^i(t,\vec{x}) \,,
\\[1ex]
\delta N = \zeta^k \partial_k N + f \dot{N} + \dot{f} N \,,
\\[1ex]
\delta N_i = \zeta^k \partial_k N_i + N_k \partial_i \zeta^k 
            + \dot{\zeta}^j g_{ij} + f \dot{N}_i + \dot{f} N_i \,,
\\[1ex]
\delta g_{ij} = \zeta^k \partial_k g_{ij} + 2 g_{k(i} \partial_{j)} \zeta^k 
                + f \dot{g}_{ij}  \,.
\end{array}                
\label{fdiff}
\end{equation}
Technically, the difference with the group of general diffeomorphisms used in GR is that the function $f$ determining the transformation of the time is not allowed to depend on the spatial coordinates. That is, under the foliation-preserving diffeomorphisms the time coordinate transforms at most into itself, which keeps the foliation. 

The complete nonprojectable theory was formulated in Refs.~\cite{Horava:2009uw,Blas:2009qj}, where the details of the formulation can be found. The main idea is to insert into the potential all the terms that preserve the gauge symmetry declared as the fundamental one, which is the foliation-preserving-diffeomorphisms symmetry, up to a given order in derivatives \cite{Blas:2009qj}. For the renormalization of the theory it is important to include terms of, at least, sixth order in spatial derivatives \cite{Horava:2009uw}. But for solutions which are of interest for large-distance applications one can deal with lower-order effective actions, since the influence of the higher order terms on these configurations can be neglected. This procedure simplifies the computations enormously. When dealing with effective actions all the terms of the given order and below that preserve the symmetry must be included. Following the spirit of effective theories, each term that is separately covariant under the gauge symmetry is written with an independent coupling constant. For example, the kinetic part has two terms that are separately covariant, $K_{ij} K^{ij}$ and $K^2$, where $K_{ij}$ is the extrinsic curvature tensor for each spatial slide of the foliation,
\begin{equation}
 K_{ij} = \frac{1}{2N} ( \dot{g}_{ij} - 2 \nabla_{(i} N_{j)} ),
\label{K}
\end{equation}
and $K$ is its trace, $K \equiv g^{ij} K_{ij}$. Therefore, a relative multiplicative constant between $K_{ij} K^{ij}$ and $K^2$ is needed, which is denoted by $\lambda$. The kinetic term arises in the action as
\begin{equation}
 K_{ij} K^{ij} - \lambda K^2 \,.
\label{kinetic}
\end{equation}
In principle any value of $\lambda$ is valid, unlike what happens in GR where the symmetry of full spacetime diffeomorphisms demands $\lambda = 1$ since $K_{ij} K^{ij}$ and $K^2$ are not separately covariant under this higher symmetry, they are only covariant when they are taken together. 

In the complete nonprojectable theory the value of $\lambda$ plays a crucial role in determining the number of propagating degrees of freedom. When $\lambda \neq 1/3$ the theory propagates three degrees of freedom: the two tensorial modes that are also present in GR plus one extra scalar mode; see, for example, Refs.~\cite{Blas:2009qj,Donnelly:2011df,Bellorin:2011ff}. But when $\lambda = 1/3$ the theory acquires two extra second-class constraints that eliminate the extra scalar degree, thus at $\lambda=1/3$ the theory propagates exactly the same degrees of freedom of GR \cite{Bellorin:2013zbp}. Interestingly, at $\lambda = 1/3$ the full kinetic term (\ref{kinetic}) acquires a conformal covariance \cite{Horava:2009uw}, but in general the terms in the potential break this conformal symmetry.

The most general effective action for large distances (second order in derivatives) with cosmological constant is \cite{Horava:2009uw,Blas:2009qj}
\begin{equation}
 S = \int dt d^3x \sqrt{g} N (  K_{ij} K^{ij} - \lambda K^2 - 2\Lambda + \xi R + \alpha a_i a^i ) \,,
\label{action}
\end{equation}
where $R$ is the spatial Ricci scalar, $a_i = \partial_i \ln N$ and $\lambda$, $\Lambda$, $\xi$ and $\alpha$ play the role of the independent coupling constants we have mentioned. $\Lambda$ is identified as the cosmological constant by analogy to GR. In (\ref{action}) we have fixed the globally multiplicative constant of the action equal to one, since such a constant does not affect the vacuum field equations. For a similar reason we may fix the value of one more coupling constant since, under the conditions we are going to impose on the field equations, one constant can be absorbed with rescalings of the other ones.

Now we start with studying the solutions of the action (\ref{action}) we are interested in. By the standard procedure of setting spherical coordinates and adapting the time coordinate to the timelike Killing vector, the most general \emph{static spherically symmetric spacetime metric} can be written in time$\times$spherical coordinates as 
\begin{equation}
ds^2 = - ( N(r)^2 - N_r N^r ) dt^2 + 2 N_r dr dt 
       + \frac{dr^2}{f(r)} + r^2 d\Omega_{(2)}^2 \,.
\end{equation}
$N_r = N_r(r)$ is the component of the shift function that is left undetermined when only the conditions of staticity and spherical symmetry are imposed and not coordinate transformation has been yet invoked. As it is well known, in GR $N_r$ can be absorbed by a coordinate transformation that implies mixing time with space. However, in Ho\v{r}ava theory such a transformation is not valid since it implies changing the foliation of the spacetime. Thus, the shift function remains as an active functional degree of freedom that distinguishes between inequivalent configurations among the set of static spherically symmetric configurations. For the sake of simplicity we shall set the shift function equal to zero as part of our ansatz, thus reducing the space of solutions. Note from (\ref{K}) that staticity and vanishing of the shift function imply $K_{ij} = 0$, and as a consequence of this $\lambda$ disappears from the field equations. Hence all the analysis is valid for the $\lambda = 1/3$ and $\lambda \neq 1/3$ theories. The vanishing of all the kinetic terms leads to a further simplification in the set of coupling constant: one of the three remaining constants $\Lambda$, $\xi$ or $\alpha$, can be absorbed by rescaling the other two, as we anticipated. From now on we set $\xi = 1$. Therefore, we have that the vacuum field equations depend on two coupling constants, $\Lambda$ and $\alpha$, under the conditions we have set. 
 
From the action (\ref{action}) we derive the field equations and then evaluate them on static configurations with vanishing shift function. This yields the equations
\begin{eqnarray}
 R^{ij} - \frac{1}{2} g^{ij} R + \Lambda g^{ij}
 - N^{-1}( \nabla^i \nabla^j N - g^{ij} \nabla^2 N )
\hspace{3em} \nonumber \\
 + \alpha N^{-2} ( \nabla^i N \nabla^j N 
    - \frac{1}{2} g^{ij} \nabla_k N \nabla^k N )
& = & 0 \,, \hspace*{3em}
\label{eisnteinstatic}
\\
 R - 2 \Lambda - \alpha ( 2 N^{-1} \nabla^2 N - N^{-2} \nabla_i N \nabla^i N ) 
& = & 0 \,.
\label{hamiltoniancstatic}
\end{eqnarray}
If $\alpha = 0$ these equations coincide with the Einstein field equations with cosmological constant in the ADM formulation under the conditions of staticity and $N_i = 0$. Equation (\ref{hamiltoniancstatic}) and the trace of (\ref{eisnteinstatic}) imply
\begin{equation}
 ( 2 - \alpha ) N^{-1} \nabla^2 N  + 2 \Lambda  =  0 \,.
\label{eqtraceprev2}
\end{equation}
Thus, field equations admit no solution for $\alpha = 2$ whenever $\Lambda \neq 0$. Assuming $\alpha \neq 2$, equations (\ref{hamiltoniancstatic}) and (\ref{eqtraceprev2}) are equivalent to
\begin{eqnarray}
 \nabla^2 N + \gamma N & = & 0 \,,
 \label{nablaN}
\\ 
 R + \alpha N^{-2} \nabla_k N \nabla^k N - ( 2 - 3\alpha )\gamma &=& 0 \,,
 \label{hamiltonianfin}
\end{eqnarray}
where $\gamma \equiv 2 \Lambda / (2 - \alpha)$. After using these equations into the field equation (\ref{eisnteinstatic}), it becomes
\begin{equation}
 R^{ij} - N^{-1} \nabla^i \nabla^j N + \alpha N^{-2} \nabla^i N \nabla^j N 
 - ( 1 - \alpha )\gamma g^{ij} 
 = 0 \,.
\label{eomstaticlag}
\end{equation}
Thus, the system to be solved for static configurations with vanishing shift function is Eq.~(\ref{eomstaticlag}) together with (\ref{nablaN}) or (\ref{hamiltonianfin}).

As we have discussed the general static spherically symmetric metric with a vanishing shift function is given by
\begin{equation}
 N = N(r) 
\,,\hspace{2em} 
 ds_{(3)}^2 = {\displaystyle\frac{dr^2}{f(r)} + r^2 d\Omega_{(2)}^2} \,.
\label{ansatz}
\end{equation}
Under this ansatz Eqs. (\ref{nablaN}) and (\ref{hamiltonianfin}) yields, respectively,
\begin{eqnarray}
 ( r^2 \sqrt{f} N' )' + \gamma \frac{r^2 N}{\sqrt{f}}
 &=& 0 \,,
\label{css2}
\\
 r f' + f - 1  - \frac{\alpha}{2} r^2 f \left( \frac{ N'}{N} \right)^2 
 + (1 - 3 \alpha / 2) \gamma r^2  &=& 0 \,.
\label{css1}
\end{eqnarray}
All off-diagonal components of the equation of motion (\ref{eomstaticlag}) vanish. The $rr$ and $\theta\theta$ components become, respectively,
\begin{eqnarray}
 \frac{f'}{r f} + \frac{N''}{N} + \frac{f' N'}{2 f N} 
   - \alpha \left( \frac{N'}{N} \right)^2  + (1 - \alpha)\gamma \frac{1}{f}  &=& 0 \,,
\label{eom1}
\\
 \frac{1}{2} r f' + f - 1 + \frac{ r f N'}{N} + ( 1 - \alpha )\gamma r^2 &=& 0 \,.
\label{eom2}
\end{eqnarray}
The system of equations (\ref{css2} - \ref{eom2}) is redundant. Indeed, after doing some computations, it turns out that the first-order equations (\ref{css1}) and (\ref{eom2}) imply the second-order ones (\ref{css2}) and (\ref{eom1}). Thus, the number of independent equations matches with the number of unknowns, $N(r)$ and $f(r)$. Since Eqs. (\ref{css1}) and (\ref{eom2}) are first-order equations there are two independent integration constants in the full set of solutions. One of these constants is evidently associated to scalings of $N$, which can always be absorbed by scaling the time. Therefore, only one integration constant has physical meaning. Although Eqs.~(\ref{css1}) and (\ref{eom2}) determine the closed system of equations, we shall strongly use the Eq.~(\ref{css2}).

The ansatz (\ref{ansatz}) includes the Schwarzschild-de Sitter and Schwarzschild-anti-de Sitter spacetimes. They are given by
\begin{equation}
 N^2 = f = 1 - \frac{\Lambda}{3} r^2 - \frac{2M}{r} \,, 
\label{schwarzschildsads}
\end{equation}
where the Schwarzschild-de Sitter metric holds when $\Lambda > 0$ and Schwarzschild-anti-de Sitter for $\Lambda < 0$, and $M$ is a free constant. If we set the GR value $\alpha = 0$, which implies $\gamma = \Lambda$, these metrics solve exactly the Eqs.~(\ref{css1}) and (\ref{eom2}) for any $\Lambda$ and $M$.


\subsection{Deformed AdS asymptotia}
Before entering in the full solutions, we study the potential asymptotic divergences for the case when the cosmological constant is negative. This helps to contrast with the anti-de Sitter metric, which is (\ref{schwarzschildsads}) with $\Lambda < 0$ and $M = 0$. To achieve this we assume that if $N$ and $f$ diverge at $r\rightarrow\infty$, then they do so with some dominant powers of $r$. That is, we assume that as $r\rightarrow\infty$,
\begin{equation}
 N = r^{a} \,,
\hspace{2em}
 f = C r^{b} \,,
\label{nfdiverging}
\end{equation}
with $a,b,C>0$ (again, Eqs.~(\ref{css1}) and (\ref{eom2}) give no information about multiplicative constants of $N$). This assumption is supported by the numerical solution, as we shall see. We then evaluate the field equations at the limit $r\rightarrow\infty$ by substituting (\ref{nfdiverging}) into them and neglecting any finite term. Equation (\ref{css2}) yields
\begin{equation}
 C a (1 + a + b/2) r^b + \gamma r^2 = 0 \,.
\end{equation}
Since none of these coefficients can be put equal to zero, we have that this equation necessarily implies
\begin{equation}
 b = 2 
\hspace{2em} \mbox{and} \hspace{2em}
 C a (2 + a) = - \gamma \,.
\label{eqac1}
\end{equation}
Next, Eq.~(\ref{eom2}) yields
\begin{equation}
 C (1 + a + b/2) r^b +  \gamma (1-\alpha) r^2 = 0 \,.
\end{equation}
With the same argument we have again that $b = 2$ and
\begin{equation}
 C (2 + a) = - \gamma (1-\alpha) \,.
\label{eqac2}
\end{equation}
We can solve $a$ and $C$ from (\ref{eqac1}) and (\ref{eqac2}), which yields
\begin{equation}
 a = \frac{1}{1 - \alpha} \,,
\hspace{2em}
 C = 
 - \frac{\Lambda}{3} 
    \left(\frac{(1 - \alpha)^2}{(1 - \alpha/2)(1 - 2\alpha/3)}\right)
\,.
\label{ac}
\end{equation}
Finally, it can be checked that with the values for $a,b,C$ given in (\ref{eqac1}) and (\ref{ac}) the Eq.~(\ref{css1}) is solved asymptotically. Thus, we see that the solutions of the field equations that diverge asymptotically with some powers of $r$ do not tend exactly to the anti-de Sitter space, but rather to a deformation of it (for small $\alpha$). Specifically, $f$ does diverge as in AdS, $f\sim r^2$, but $N$ exhibits a slight modification, $N^2 \sim r^{2(1-\alpha)^{-1}}$, which for small $\alpha$ can be seen as a deformation of $r^2$. The field equations fix the coefficient of the dominant mode of $f$, which is the constant $C$ given in (\ref{ac}) (actually, this coefficient is also different to the one of AdS). Thus, the dominant mode of $f$ at infinity is completely fixed for all solutions 


\subsection{The coordinate transformation}
Let us start our approach for solving the system of equations. We remark that in the whole procedure we shall assume that $\alpha$ is near to zero. There is a quadratic structure encoded in Eqs.~(\ref{css1}) and (\ref{eom2}). Indeed, a combination of these two equations yields the quadratic equation
\begin{equation}
\frac{f}{\rho^2} \left[ 
 \left( 1 + \frac{r N'}{N} \right)^2
 - \left( \frac{ \beta r N'}{N} \right)^2 \right] = 1 \,,
\label{quadraticgeneral}
\end{equation}
where
\begin{equation}
 \rho^2 \equiv 1 - \Lambda r^2 \,,
\hspace{2em}
 \beta \equiv \sqrt{1 - \alpha / 2} \,.
\end{equation}
Our assumption for $\alpha$ implies that $\beta$ is close to one. This quadratic equation suggests the implementation of a procedure similar to the one we used in the $\Lambda=0$ case \cite{Bellorin:2014qca}. 

To further proceed we choose a negative cosmological constant from now on. With $\Lambda < 0$ we have that $\rho^2 > 0$ everywhere, so the analysis of Eq.~(\ref{quadraticgeneral}) simplifies greatly. Under this condition we may write the Eq.~(\ref{quadraticgeneral}) as
\begin{equation}
 \left( \frac{\sqrt{f}}{\rho} + \frac{r \sqrt{f} N'}{\rho N} \right)^2
 - \left( \frac{ \beta r \sqrt{f} N'}{\rho N} \right)^2 = 1 \,. 
\label{quadratic}
\end{equation}
The general solution to this equation can be written as
\begin{equation}
 \frac{ \beta r \sqrt{f} N'}{\rho N} = \sinh{\chi}  \,,
\hspace{2em}
 \frac{\sqrt{f}}{\rho} + \frac{r \sqrt{f} N'}{\rho N} = \pm \cosh{\chi} \,,
\label{particularsolution}
\end{equation}
for any $\chi \in (-\infty,+\infty)$. Thus, we have that in principle there are two sets of solutions, each one corresponding to a choice of sign in (\ref{particularsolution}). It turns out that the two sets of geometries are equivalent after coordinate transformations. We shall further comment on this later on. Thus, we may take (\ref{particularsolution}) with the upper sign as the general solution of Eq.~(\ref{quadratic}) without loss of generality. Solution (\ref{particularsolution}) is then equivalent to
\begin{eqnarray}
 \frac{\sqrt{f}}{\rho} &=& 
   \cosh{\chi} - \beta^{-1} \sinh{\chi}  \,,
\label{sqrtf}
\\
 \frac{r N'}{N} &=& 
 \frac{1}{\beta} \left( \frac{\cosh{\chi}}{\sinh{\chi}} - \frac{1}{\beta} \right)^{-1} \,.
\label{nprime}
\end{eqnarray}

Now our aim is to use $\chi$ as a new radial coordinate, obtaining, first of all, an equation for the coordinate transformation between $r$ and $\chi$ and then equations for $N$ and $f$ in terms of $\chi$. For this goal we join Eq.~(\ref{css2}) with the Eqs.~(\ref{sqrtf}) and (\ref{nprime}) to form  a system of equations for the functions $r(\chi)$, $N(\chi)$ and $f(\chi)$. Notice that in this procedure we are replacing the original field equations (\ref{css1}) and (\ref{eom2}) with the Eqs.~(\ref{css2}) and (\ref{quadratic}), which form an equivalent system of two field equations.

We first rewrite the differential equation (\ref{css2}) as an equation with $\chi$ as the independent variable, obtaining
\begin{equation}
 \frac{d}{d\chi} \left( r^2 \sqrt{f} \frac{dN}{dr} \right)
 + \gamma \frac{r^2 N}{\sqrt{f}} \frac{dr}{d\chi} = 0 \,.
\end{equation}
This equation can be brought to the form
\begin{equation}
 \left(\frac{1}{N} \frac{dN}{d\chi} \right)
 r^2 \rho \left( \frac{\sqrt{f}}{\rho}\right) \left( \frac{1}{N} \frac{dN}{dr} \right)
 + \frac{d}{d\chi} \left( r^2 \rho \left(\frac{\sqrt{f}}{\rho}\right) \left( \frac{1}{N} \frac{dN}{dr} \right) \right)
 + \gamma \left( \frac{\rho}{\sqrt{f}} \right) \frac{r^2}{\rho} \frac{dr}{d\chi} = 0 \,,
\end{equation}
and finally to the form
\begin{equation}
\begin{array}{rcl}{\displaystyle
\left[ 
\left( \frac{\sqrt{f}}{\rho} \right)^2 \left( \frac{r}{N}\frac{dN}{dr} \right)^2 \rho^2
+ \left( \frac{\sqrt{f}}{\rho} \right)^2 \left( \frac{r}{N}\frac{dN}{dr} \right)
     ( 1 - 2 \Lambda r^2 ) 
+ \gamma r^2 \right] \frac{dr}{d\chi} = } \hspace*{2em} &&
\\ {\displaystyle 
- \left( \frac{\sqrt{f}}{\rho} \right) \frac{d}{d\chi} 
  \left[ \left( \frac{\sqrt{f}}{\rho} \right) \left( \frac{r}{N}\frac{dN}{dr} \right) \right] r \rho^2 \,. } &&
\end{array}
\end{equation}
By substituting Eqs.~(\ref{sqrtf}) and (\ref{nprime}) into this last equation and after a bit of algebra we obtain a differential equation only for $r(\chi)$,
\begin{equation}
  \frac{dr}{d\chi} 
 = 
 \left( \frac{1}{\beta} - \frac{\cosh{\chi}}{\sinh{\chi}} \right) 
    \frac{r \rho^2}{H} \,,
\label{diffcoordtransf}
\end{equation}
where
\begin{equation}
 H \equiv 1 - \Lambda \left( 2 - \frac{\cosh{\chi}}{\beta \sinh{\chi}}  \right) r^2 \,. 
\label{H}
\end{equation}
Equation (\ref{diffcoordtransf}) determines the coordinate transformation from $\chi$ to $r$.

The space-time metric can be expressed in the $\chi$ coordinate by using Eqs.~(\ref{sqrtf}) and (\ref{diffcoordtransf}), which yields
\begin{equation}
 ds^2 = - N(\chi)^2 dt^2 +  h(\chi) d\chi^2 + r(\chi)^2 d\Omega^2_{(2)} \,,
 \label{metricchi}
\end{equation}
where
\begin{equation}
 h(\chi) = \frac{r^2 \rho^2}{\sinh^2\chi H^2} \,.
 \label{hcomponent}
\end{equation}
Thus, the radial component $h(\chi)$ is algebraically determined once $r(\chi)$ has been obtained. Now we use the chain rule in Eq.~(\ref{nprime}) and substitute Eq.~(\ref{diffcoordtransf}) into it, obtaining
\begin{equation}
 \frac{1}{N}\frac{dN}{d\chi} = - \frac{\rho^2}{\beta H} \,.
\label{dNdchi}
\end{equation}
This equation allows to find $N(\chi)$ once $r(\chi)$ has been found from (\ref{diffcoordtransf}). Under this approach we have that the integration of Eqs.~(\ref{diffcoordtransf}) and (\ref{dNdchi}) leads to the two independent integration constants we discussed above.  Finally, we have that the solutions of the field equations are determined by Eqs.~(\ref{diffcoordtransf}), (\ref{hcomponent}) and (\ref{dNdchi}).

We recall the reader that, to arrive at this point, in Eqs.~(\ref{particularsolution}) we chose the solution with plus sign. When the minus sign is taken the three equations determining the solutions are modified, but the reparameterization $\chi' = -\chi$ brings them exactly to the form in (\ref{diffcoordtransf}), (\ref{dNdchi}) and (\ref{hcomponent}). Therefore, any choice of sign in Eqs.~(\ref{particularsolution}) leads to the same configurations modulo coordinate transformations.

We comment that at the point $\Lambda = 0$ Eq.~(\ref{diffcoordtransf}) becomes
\begin{equation}
\frac{dr}{d\chi}  = 
 \left( \frac{1}{\beta} - \frac{\cosh{\chi}}{\sinh{\chi}} \right) r \,,
\label{minkowskian}
\end{equation}
and its integral is
\begin{equation}
 r(\chi) = \frac{ k e^{\chi/\beta}}{\sinh{\chi}} \,.
\label{rlambdacero}
\end{equation}
This is the coordinate transformation we found for the case without cosmological constant \cite{Bellorin:2014qca}.


\subsection{Analysis of the field equations}
Having found the differential equations that determines $r(\chi)$, $N(\chi)$ and $h(\chi)$, the rest of our procedure will consist of in doing analysis to extract special points of these equations and then performing numerical integration on them. For the integration the key equation is (\ref{diffcoordtransf}), since it is a single equation for $r(\chi)$ and the solutions for $N$ and $h$ depend on it. In turn, for the numerical solution of Eq.~(\ref{diffcoordtransf}) it is mandatory to analyze the several singularities this equation has as well as its critical points. From now on we assume a positive $\alpha$, such that $\alpha \sim 0^+$ and $\beta \sim 1^-$. In the Appendix we summarize the results for negative $\alpha$.

We start with the singularities. It is easy to see that at both asymptotic limits, $\chi = \pm \infty$, Eq.~(\ref{diffcoordtransf}) implies that $dr/d\chi$ is positive, hence $r$ becomes monotonically increasing at both ends of $\chi$. Thus, there is a divergence at $\chi = +\infty$ that yields $r= \infty$, the spatial infinity. On the other hand, as $\chi \rightarrow -\infty$ the radius $r$ necessarily tends to zero. We shall check this behavior at both asymptotic limits it in the numerical integration. We remark that at the divergence $\chi=+\infty$ the factor $H$ diverges. 

Further divergences of $dr/d\chi$ arise for finite $\chi$ (with no analogs in the $\Lambda = 0$ case). The key feature to localize these divergences is that \emph{they are just the zeroes of $H$}. Indeed, another potential source of divergences is the factor 
\begin{equation}
\left( \frac{1}{\beta} - \frac{\cosh{\chi}}{\sinh{\chi}} \right) \,,
\label{firstfactor}                                        
\end{equation}
which diverges at $\chi = 0$, but it cancels out with the divergence of $H$ at the same point. Note that this behavior is completely different to the singularity of the ``Minkowskian'' equation (\ref{minkowskian}), which effectively diverges at $\chi = 0$. Therefore, characterizing the singularities of Eq.~(\ref{diffcoordtransf}) for finite $\chi$ is equivalent to study the zeroes $H$. 

We find that there are two kinds of zeroes of $H$ depending on whether $r$ diverges or not:
\begin{enumerate}
\item {\bf Singularities of the first kind} These are zeroes of $H$ at which $r$ diverges. Let us denote by $\chi^{(1)}_s$ a zero of $H$ of this kind. It turns out that the factor
\begin{equation}
  \left( 2 - \frac{\cosh{\chi}}{\beta \sinh{\chi}}  \right)
\label{factorchi}
\end{equation}
arising in $H$ must necessarily vanish at $\chi^{(1)}_s$ in order to compensate the divergence of the $r^2$ factor that multiplies it, such that $H$ remains finite. Therefore, a zero $\chi^{(1)}_s$ of $H$ is necessarily a zero of the factor (\ref{factorchi}). Since this factor does not depend on $r$ its roots are fixed for all solutions, 
\begin{equation}
  \tanh{\chi^{(1)}_s} = \frac{1}{2\beta} \,.
\label{singularity1}
\end{equation}
This has only one solution for any given $\beta$, so there is only one $\chi_s^{(1)}$. With fixed we mean that this value is independent of any initial condition corresponding to any particular solution $r(\chi)$; all the solutions $r(\chi)$ that have this singularity reach it at the same point (\ref{singularity1}). Note that the converse for this behavior is not true in general: the factor (\ref{factorchi}) can be zero, but $r$ may remain finite and in this case necessarily $H=1$, so there is no divergence. That is, in the full set of solutions reaching the point (\ref{singularity1}), not necessarily all of them develop a divergence, some may pass through this point regularly.

Since in this kind of divergence one has $r=\infty$, it corresponds to the spatial infinity.

\item {\bf Singularities of the second kind} In the other kind of zeroes of $H$, denoted by $\chi^{(2)}_s$, $r$ remains finite. In this case the factor (\ref{factorchi}) cannot vanish. Instead, one may obtain the value of $r$ from Eq (\ref{H}), 
\begin{equation}
 -\Lambda (r^{(2)}_s)^2 =  
 \left( \frac{\cosh{\chi^{(2)}_s}}{\beta \sinh{\chi^{(2)}_s}} 
    - 2 \right)^{-1} \,. 
\label{secondkind}
\end{equation}
The value of $\chi_s^{(2)}$ is not fixed in the set of solutions and the pair $(\chi^{(2)}_s,r^{(2)}_s)$ corresponding to this kind of zeroes varies among each solution. Since relation (\ref{secondkind}) fixes the value of $r$ in terms of the one of $\chi$, solutions passing through some $\chi_s^{(2)}$ may or may not exhibit the singularity depending on whether the image $r(\chi_s^{(2)})$ coincides or not with $r_s^{(2)}$. In the last case $H$ is nonvanishing. Thus, we have again that some solutions can posses this singularities whereas others do not.

\end{enumerate}

It comes out the question whether $\chi_s^{(1)}$ and $\chi_s^{(2)}$ are just failures of the $\chi$ coordinate, hence coordinate singularities, or essential singularities. In the case of $\chi_s^{(1)}$ the asymptotic limit $r\rightarrow\infty$ is reached, hence there is a place for having at $\chi\rightarrow\chi_s^{(1)}$ the deformed AdS asymptotia we have discussed. If this is the case $\chi_s^{(1)}$ would be a coordinate singularity and the divergence of the coordinate transformation (\ref{diffcoordtransf}) at $\chi_s^{(1)}$ would signal that we are reaching at one boundary of the space that is regular. In the case of $\chi_s^{(2)}$ we have that $r$ acquires a finite value. If $\chi_s^{(2)}$ is also a coordinate singularity, such that the final solution is regular at $\chi_s^{(2)}$, then the solution written in terms of $\chi$ will stop at $r_s^{(2)}(\chi_s^{(2)})$ but it could be continued if there are other available solutions ending regularly at the same value of $r_s^{(2)}$. We shall investigate these issues with the numerical evaluation of the metric components and their derivatives as well as the curvature.

So far we have seen that the spatial infinity $r = \infty$ is reached in the $\chi$ coordinate at $\chi = +\infty$ and $\chi = \chi_s^{(1)}$. Are there other spatial infinities in the coordinate $\chi$?. We may find analytically the asymptotic solution to the coordinate transformation (\ref{diffcoordtransf}). With it we may proof rigorously that the spatial infinity, $r = \infty$, is reached in the coordinate $\chi$ \emph{only} at $\chi =  +\infty$ and $\chi = \chi^{(1)}_s$. This information is crucial to understand the behavior of the numerical solutions. To achieve this we first rewrite the Eq.~(\ref{diffcoordtransf}) in the form
\begin{equation}
 \frac{dr}{d\chi}  = 
 \frac{ \beta^{-1} - \tanh{\chi} }
  {1 - {\Lambda r^2}{\rho^{-2}} \left( 1 - \beta^{-1} \tanh{\chi} \right)} r   \,,
\label{diffcoordtransfalternative}
\end{equation}
and then consider the asymptotic limit $r \rightarrow \infty$ into it. $\rho^2$ can be substituted by $\rho^2 = - \Lambda r^2$, such that in the asymptotic limit the Eq.~(\ref{diffcoordtransfalternative}) becomes
\begin{equation}
 \frac{dr}{d\chi}  = 
 \frac{ {\beta}^{-1} - \tanh{\chi} }
  { 2 - \beta^{-1} \tanh{\chi}} r \,.
\label{diffcoordtransfasymp}
\end{equation}
This equation can be integrated directly. The result is
\begin{equation}
 r = 
 \frac{ k ( 1 + \tanh{\chi} )^m}
  {|\tanh{\chi^{(1)}_s} - \tanh{\chi}|^n (1 - \tanh{\chi} )^{\tilde{n}}} \,,
\label{rinfinity}
\end{equation}
where $k$ is an integration constant and
\begin{equation}
 m = \frac{1 + \beta}{4\beta + 2} \,,
\hspace{2em}
 n = \frac{ 2\beta^2 - 1 }{4\beta^2 - 1} \,,
\hspace{2em}
 \tilde{n} = \frac{ 1 - \beta}{4\beta - 2} \,.
\end{equation}
For $\beta \sim 1^{-}$ ($\alpha \sim 0^{+}$) we have $m,n,\tilde{n} > 0$. Therefore, Eq.~(\ref{rinfinity}) implies that $r = \infty$ is reached only at $\chi = \chi^{(1)}_s$ and $\chi = +\infty$. Moreover, one may estimate the error of the asymptotic solution (\ref{rinfinity}) by comparing the equation it satisfies with the original equation. At $r\rightarrow\infty$, the right hand sides of Eqs.~(\ref{diffcoordtransfalternative}) and (\ref{diffcoordtransfasymp}) differ by terms of order $1/r^3$ and lower. Thus, the error of the asymptotic solution (\ref{rinfinity}) is of order $1/r^3$.

Now we move to the critical points of the solutions of Eq.~(\ref{diffcoordtransf}). For general (negative) $\Lambda$, equation (\ref{diffcoordtransf}) implies that $r(\chi)$ has a finite critical point $\hat{\chi}$ at the point where the factor (\ref{firstfactor}) vanishes; that is, at
\begin{equation}
 \tanh{\hat{\chi}} = \beta \,.
\label{throat}
\end{equation}
Since this condition is independent of $r$, we have that this critical point is also fixed. Surprisingly, this is the same location of the throat in the corresponding $\chi$ coordinate for the $\Lambda = 0$ case \cite{Bellorin:2014qca}. In the numerical solution we are going to see that $\hat{\chi}$ is effectively a critical point since $r$ remains finite ($r \rho^2$ is finite) and $H$ does not vanish at $\hat{\chi}$. As for the $\chi_s^{(1)}$ and $\chi_s^{(2)}$ singularities, there remains to elucidate whether $\hat{\chi}$ is a regular point of the solution, such that  $\hat{\chi}$ just labels a regular point at which the $r$ coordinate fails. This will be checked with the numerical evaluation of the metric components and their derivatives in the $\chi$ coordinate. Taking a derivative to Eq.~(\ref{diffcoordtransf}) and evaluating at $\hat{\chi}$ we get
\begin{equation}
 \left[\left( 1 - \gamma ( 1 - \alpha ) r^2 \right)
  \frac{d^2 r}{d\chi^2} \right]_{\hat{\chi}} =
 \sinh^{-2}{\hat{\chi}} \left( r - \Lambda r^3 \right)_{\hat{\chi}} \,.
\label{seconderivative}
\end{equation}
Recalling that we are considering a small $\alpha$ and a negative $\Lambda$, Eq.~(\ref{seconderivative}) implies that the second derivative of $r(\chi)$ is positive at $\hat{\chi}$. Therefore, $\hat{\chi}$ is a minimum of $r(\chi)$. This shows that any solution possessing the point $\hat{\chi}$ has a throat at $\hat{\chi}$.\footnote{Note that this conclusion is based only on the unavoidable presence of a minimum of $r(\chi)$ at the fixed value (\ref{throat}), with no assumption on the value of $r$ itself, unlike the behavior of the singular points $\chi_s^{(1)}$ and $\chi_s^{(2)}$. Thus, any solution passing through $\hat{\chi}$ necessarily possesses a throat.} 

As a final analysis that will help to understand the numerical results, we study the ordering of these special points. First, by reversing the Eq.~(\ref{secondkind}), we may write it as
\begin{equation}
 \tanh{\chi^{(2)}_s} = 
  \left( 1 - \frac{1}{2\Lambda r^{(2)}_s} \right)^{-1} 
	  \tanh{\chi^{(1)}_s} \,.
\end{equation}
Since $\Lambda r_s^{(2)} < 0$ and $\chi^{(1)}_s > 0$, this relation implies $0 < \chi^{(2)}_s < \chi^{(1)}_s$ for all finite $r^{(2)}_s$. We stress that this ordering holds regardless of the fact that the value of $\chi^{(2)}_s$ is different for each solution whereas $\chi^{(1)}_s$ is fixed. Thus, $\chi^{(1)}_s$ plays the role of upper bound for all the possible values of $\chi^{(2)}_s$. Second, we recall that we are assuming $\alpha$ near to zero, which yields $\beta$ close to one. Under this condition we have that $\hat{\chi} > \chi^{(1)}_s$ (this breaks down at $\beta = 1/\sqrt{2}$, far enough from $\beta = 1$). Therefore, we find the universal ordering
\begin{equation}
 0 < \chi^{(2)}_s < \chi^{(1)}_s < \hat{\chi} \,,
\label{ordering}
\end{equation}
for $\alpha$ close to zero. Notice that no singularities are developed in the domain $(\chi_s^{(1)},+\infty)$, except for the asymptotic singularity $\chi \rightarrow +\infty$.


\subsection{Numerical solutions}

\subsubsection{Global numerical characterization}
\begin{figure}[!ht]
\begin{center}
\includegraphics[scale=0.55]{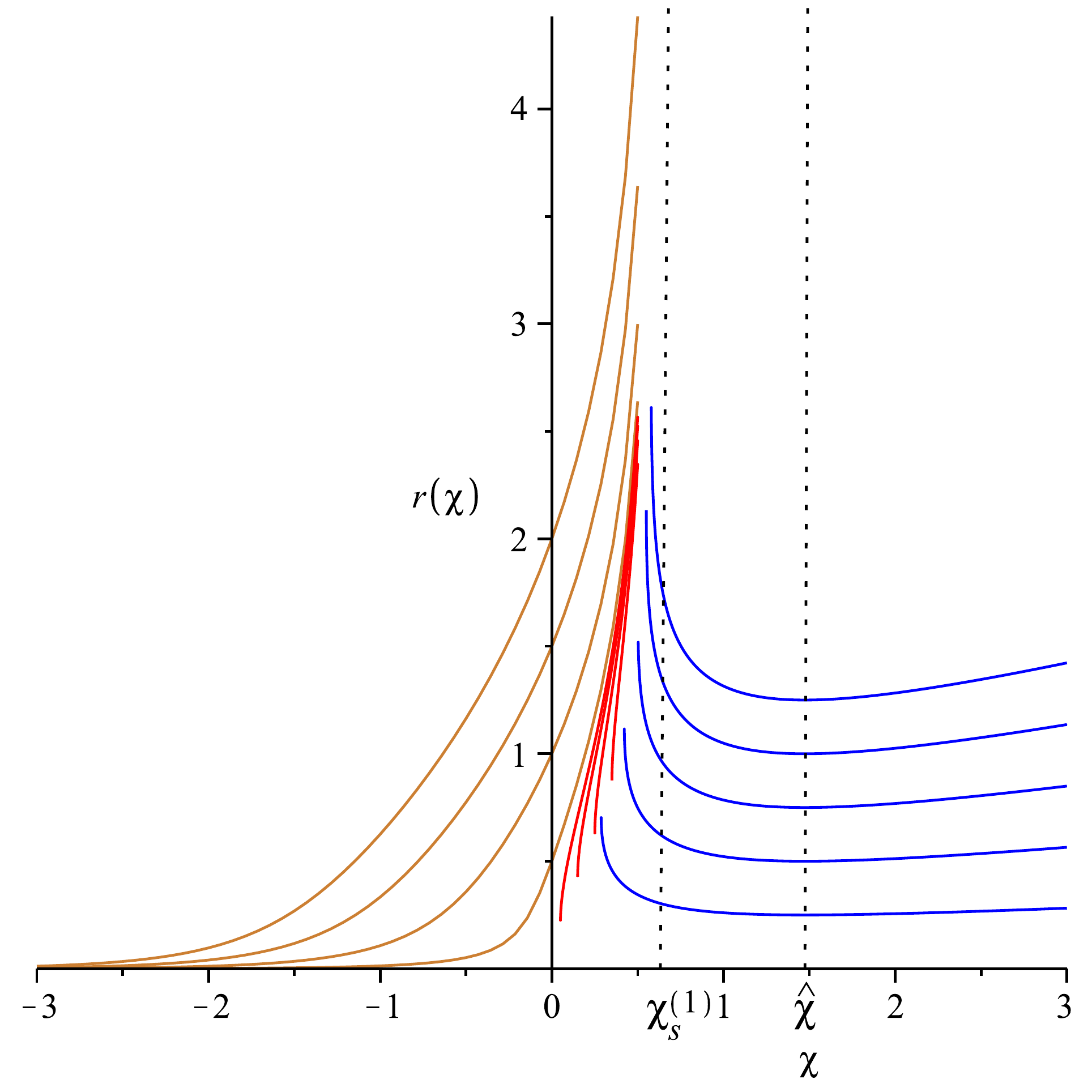}
\caption{\label{fig:allsolutions} \small The several numerical solutions $r(\chi)$ of the Eq.~(\ref{diffcoordtransf}). The dynamical constants have been set as $\beta = 0.9$ and $\Lambda = -1$, which yields the minimum approximately at $\hat{\chi} = 1.47$ and the first-kind singularity at $\chi^{(1)}_s = 0.67$. The solutions that develop the second-kind singularity $\chi_s^{(2)}$ are the blue and red curves. The $\chi_s^{(2)}$ values, which the reader can locate at the left end of each of these curves, vary among each curve and all of them fall in the subdomain $(0,\chi^{(1)}_s)$.  There are three classes of continuous solutions, each one with a definite domain in $\chi$. The  curves of class 1 (blue) start at their second kind singularity, pass through the critical point $\hat{\chi}$, where they acquire a minimum, and then grow monotonically up to spatial infinity. The domain for these solutions is $\chi \in (\chi^{(2)}_s, +\infty)$ and their images are composed of two sides. The image of $\chi \in (\hat{\chi},+\infty)$ is the side $r \in (\hat{r},\infty)$ whereas the image of $\chi \in (\chi_s^{(2)} , \hat{\chi})$ is the side $r \in (\hat{r}, r^{(2)}_s)$, where $\hat{r}$ is the image of $\hat{\chi}$. The class 2 (red) start at their second kind singularity and grow 
monotonically up to the first kind singularity where they reach the spatial infinity. They cover the domain $\chi \in (\chi^{(2)}_s,\chi^{(1)}_s)$ and range $r \in (r^{(2)}_s , \infty)$. The curves of class 3 (golden) are also monotonically increasing. They cover all the domain $\chi \in (-\infty,\chi^{(1)}_s)$ with range $r \in (0,\infty)$. Since $0 < \chi^{(2)}_s < \chi^{(1)}_s < \hat{\chi}$, the classes 1 and 2 never pass to negative $\chi$, the curves of class 2 are always located in subdomains of $(0,\chi^{(1)}_s)$ and the classes 2 and 3 do not develop the minimum $\hat{\chi}$.
}
 \end{center}
\end{figure}

Now we are ready to perform the numerical integration. Equation (\ref{diffcoordtransf}) requires the specification of one initial condition $r(\chi_0)$, where $\chi_0$ can take any value in $(-\infty,+\infty)$. Since (\ref{diffcoordtransf}) with a given $r(\chi_0)$ constitutes a closed initial-value problem, different curves $r(\chi)$ cannot intersect themselves. We give several initial conditions by varying $\chi_0$ over the subsets specified in (\ref{ordering}). In Fig.~\ref{fig:allsolutions} we present the several kinds of solutions the numerical evaluation of Eq.~(\ref{diffcoordtransf}) yields.

There are three kinds of curves:

\begin{enumerate}
\item {\bf Class 1: solutions with a minimum}
The first class of curves possesses the critical point $\hat{\chi}$, whose location is the same for all this class of solution. They can be generated by giving initial data at the right of the first-kind singularity, $\chi_0 > \chi_s^{(1)}$. As we discussed above, there is no singularity in $(\chi_s^{(1)},+\infty)$, hence all these curves pass over $\hat{\chi}$ and get a minimum there. For $\chi\rightarrow +\infty$ they grow to infinity since they tend to the asymptotic singularity of Eq.~(\ref{diffcoordtransf}). We see that at the left of $\hat{\chi}$ they pass through $\chi_s^{(1)}$ without developing any divergence, but further at the left they meet a singularity of the second kind, such that the solution is valid up to this point and $r$ remains finite there.

The numerical integration shown in Fig.~\ref{fig:allsolutions} indicates that these solutions have two sides joined by a throat. In one side (from $\chi=+\infty$ to $\chi=\hat{\chi}$) they cover the space from infinity down to the throat; in the other side they cover the space increasing from the throat up to the finite value of $r$ determined by the singularity of the second kind $\chi^{(2)}_s$. 

\item {\bf Class 2: intermediate curves}
The second class of solutions have both the singularities of the first and second kind. They can be seen in the plot as the intermediate curves that start at some finite value of $\chi$ located in $(0,\chi_s^{(1)})$ and then increase monotonically. The points where they start are singularities of the second kind, so $r$ remains finite there. These solutions are not valid for values of $\chi$ lower than their corresponding $\chi^{(2)}_s$ singularities. As $\chi$ increases from the corresponding $\chi^{(2)}_s$ the solutions grow monotonically up until they reach the singularity of the first kind $\chi^{(1)}_s$, which is universal and $r=\infty$ is reached. Thus, these solutions cover a subset of the space from the infinity down to the finite value $r^{(2)}_s$ determined by the singularity of the second kind they reach to the left. 

\item {\bf Class 3: Solutions with the origin} 
The third class of solutions are the ones located at the left of the graph. They start at $\chi \rightarrow -\infty$, where $r=0$, and grow up until they reach the singularity of the first kind at $\chi^{(1)}_s$, where $r = \infty$. They can be generated by given initial data at  $\chi_0 = 0$. These solutions cover a single copy of all the possible space in $r$. They never develop a singularity of the second kind despite of the fact that they pass over values of $\chi$ where solutions 1 and 2 reach these singularities.

\end{enumerate}

\begin{figure}[!ht]
 \begin{center}
 \includegraphics[scale=0.34]{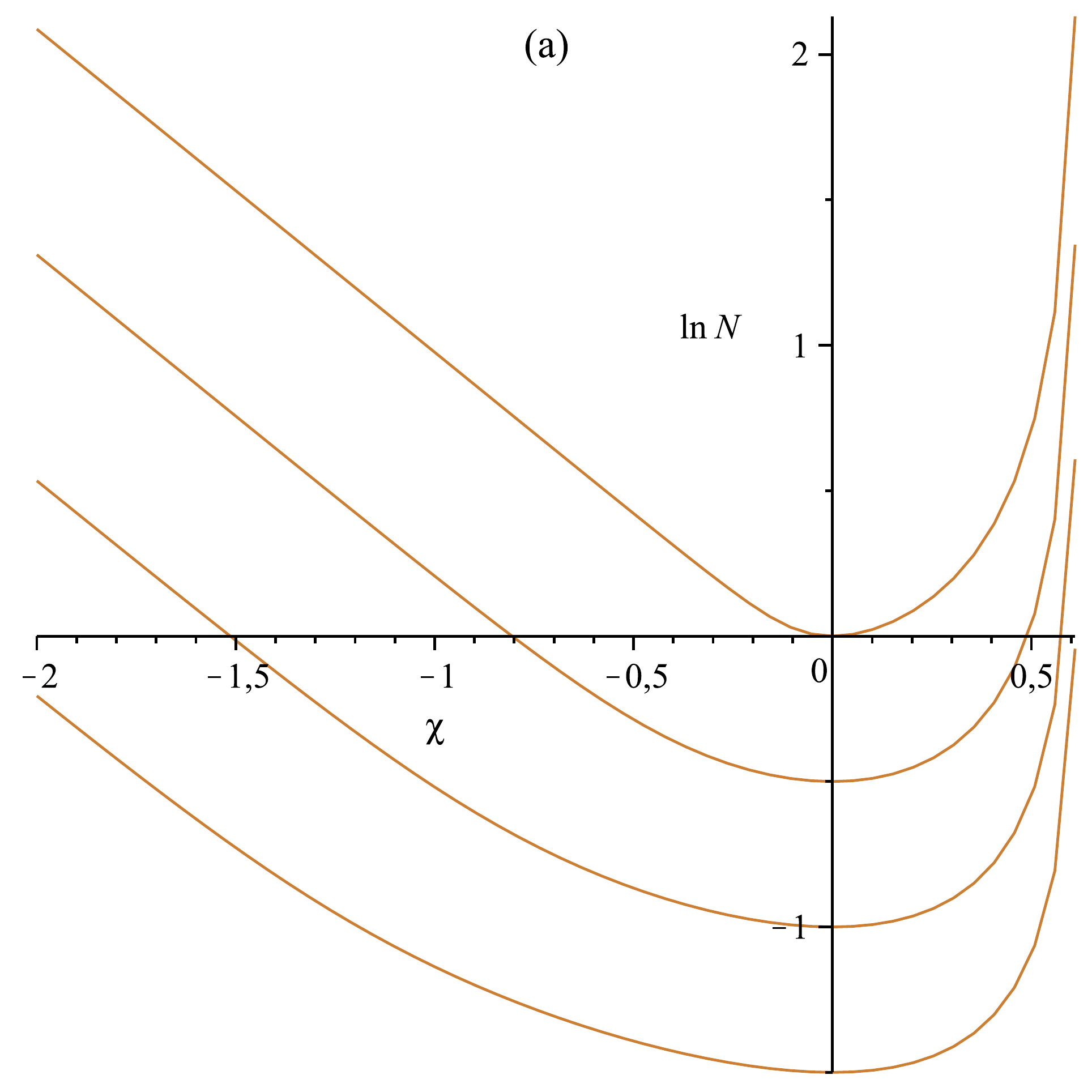}
 \includegraphics[scale=0.34]{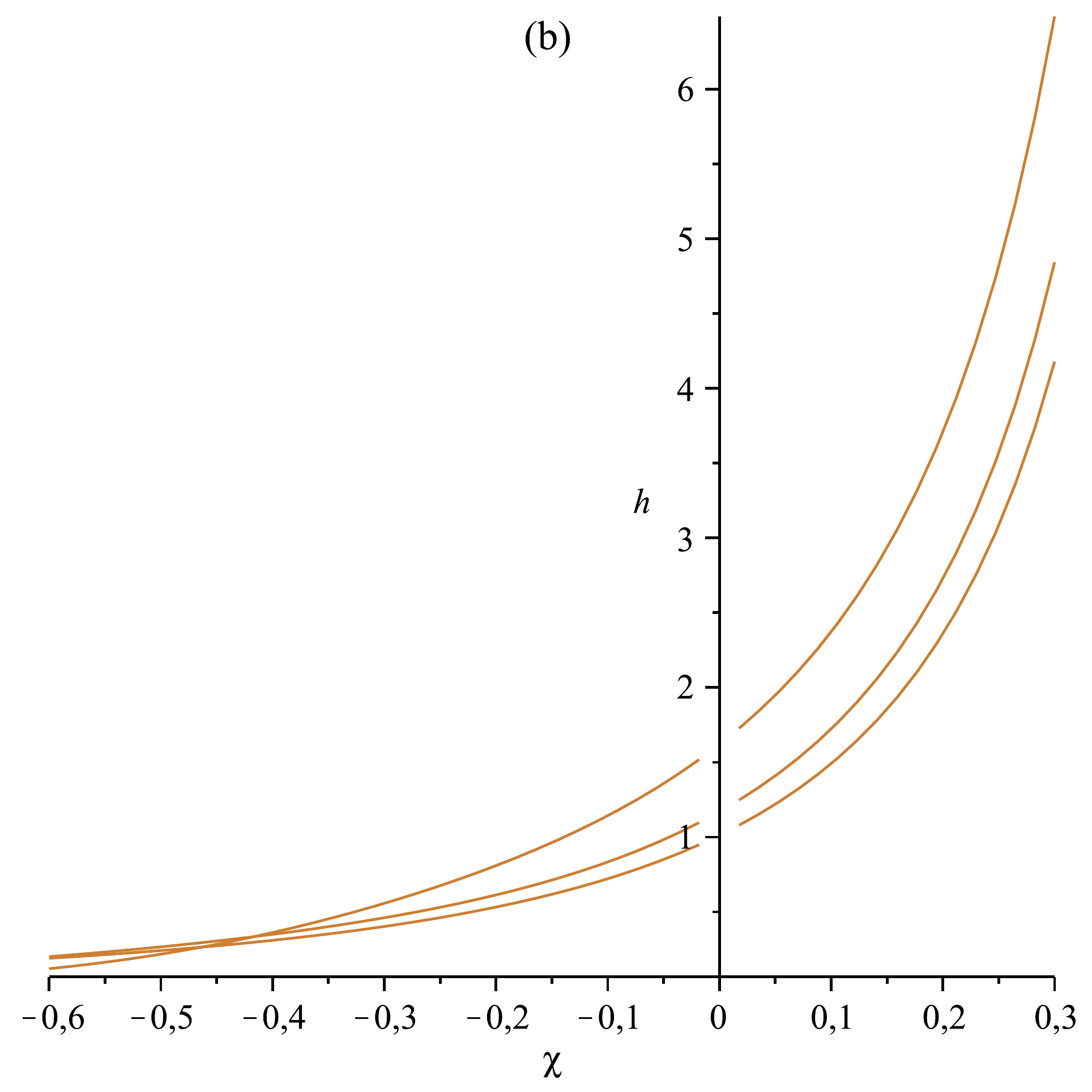}
 \caption{\label{fig:nhnaked} \small Metric components $\ln{N}$ (a) and $h$ (b) for the solution of class 3 which contains the origin. The functions $N$ and $h$ are regular in the domain $\chi\in(-\infty,\chi_s^{(1)})$ which is in one-to-one correspondence with $r\in(0,\infty)$. $N$ diverges and $h$ goes to zero at $\chi\rightarrow -\infty$ ($r\rightarrow 0$); whereas $N$ and $h$ diverge as $\chi\rightarrow\chi_s^{(1)}$ ($r\rightarrow \infty$). The values of $\beta$ and $\Lambda$ are the same of Fig.~\ref{fig:allsolutions}.}
 \end{center}
\end{figure}

To obtain the metric components we must solve the Eq.~(\ref{dNdchi}) numerically. This requires the specification of a further (and last) initial condition, $N(\chi_0)$. In Fig.~\ref{fig:nh} we show the plots for the metric components $N(\chi)$ and $h(\chi)$ for the classes of solutions 1 and 2 and in Fig.~\ref{fig:nhnaked} for the class 3. Since the Eq.~(\ref{dNdchi}) for $N(\chi)$ depends on $r(\chi)$, To a given initial condition $N(\chi_0)$ there correspond several solutions (distinguished among them by $r(\chi_0)$), hence the curves $N(\chi)$ intersect themselves in general. The same happens with $h(\chi)$ since its formula (\ref{hcomponent}) is a noninjective function of $r$. For the sake of clearness we have plotted only a representative group  of (almost) nonintersecting curves. 

The main features are:

\begin{figure}[!ht]
 \begin{center}
 \includegraphics[scale=0.34]{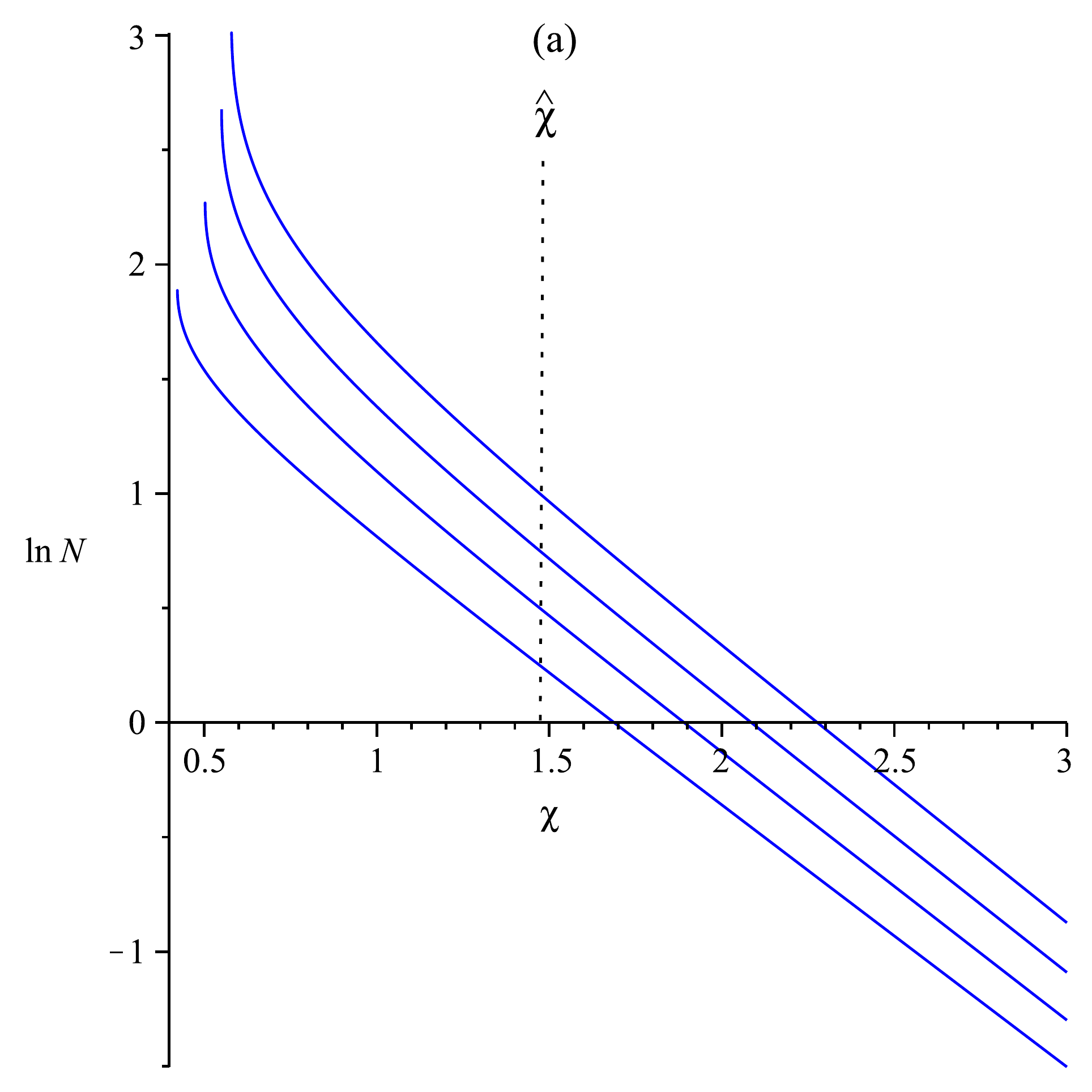}
 \includegraphics[scale=0.34]{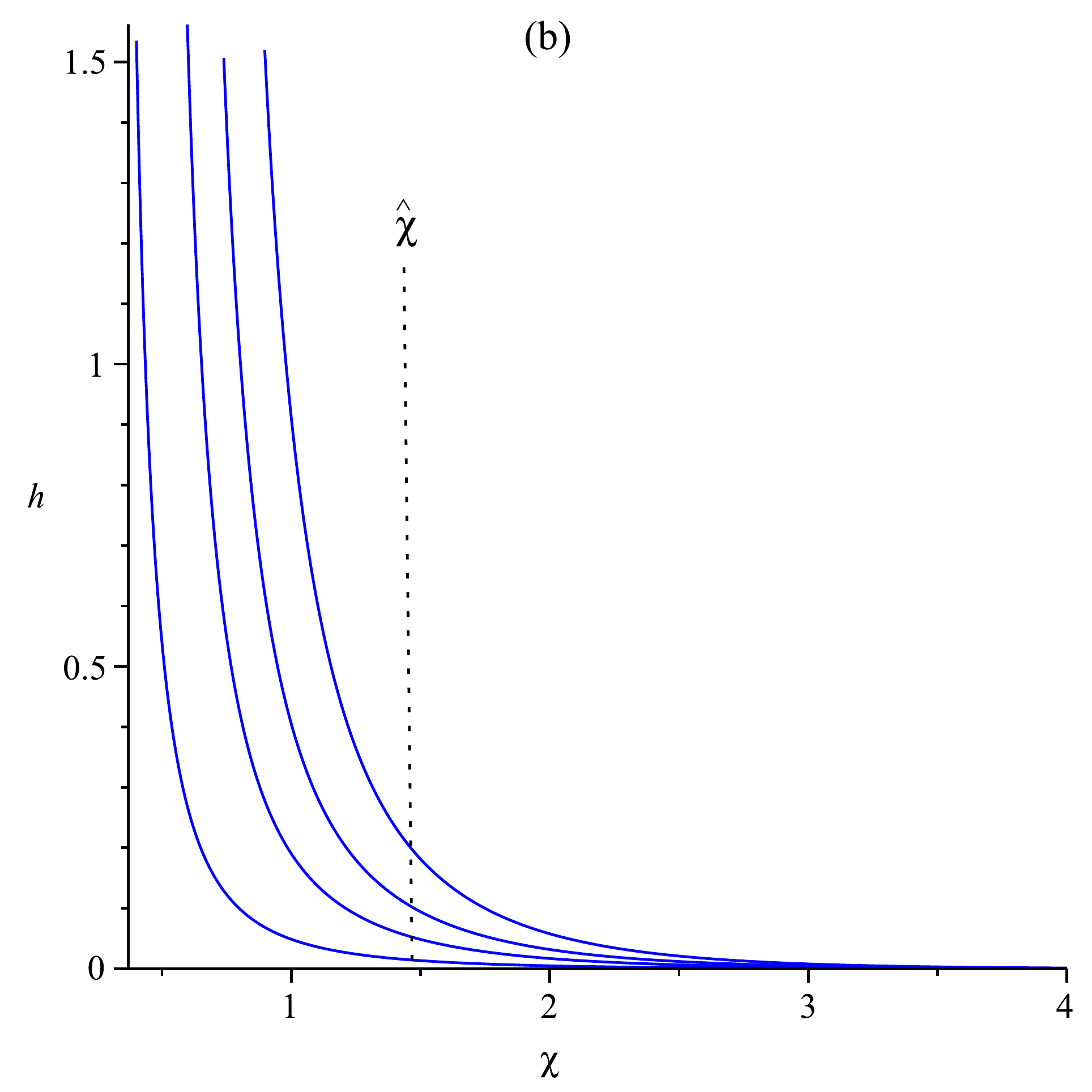}
 \includegraphics[scale=0.34]{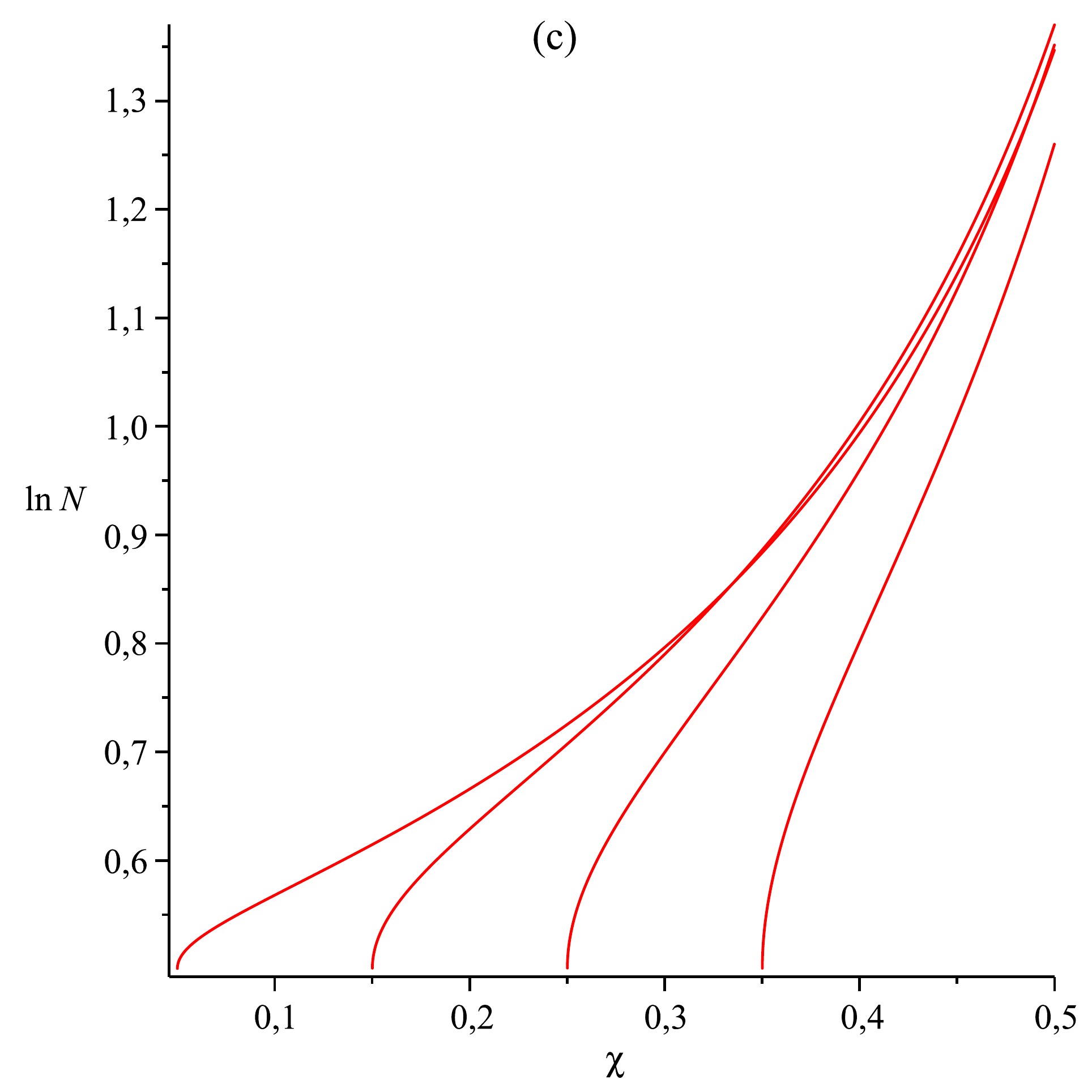}
 \includegraphics[scale=0.34]{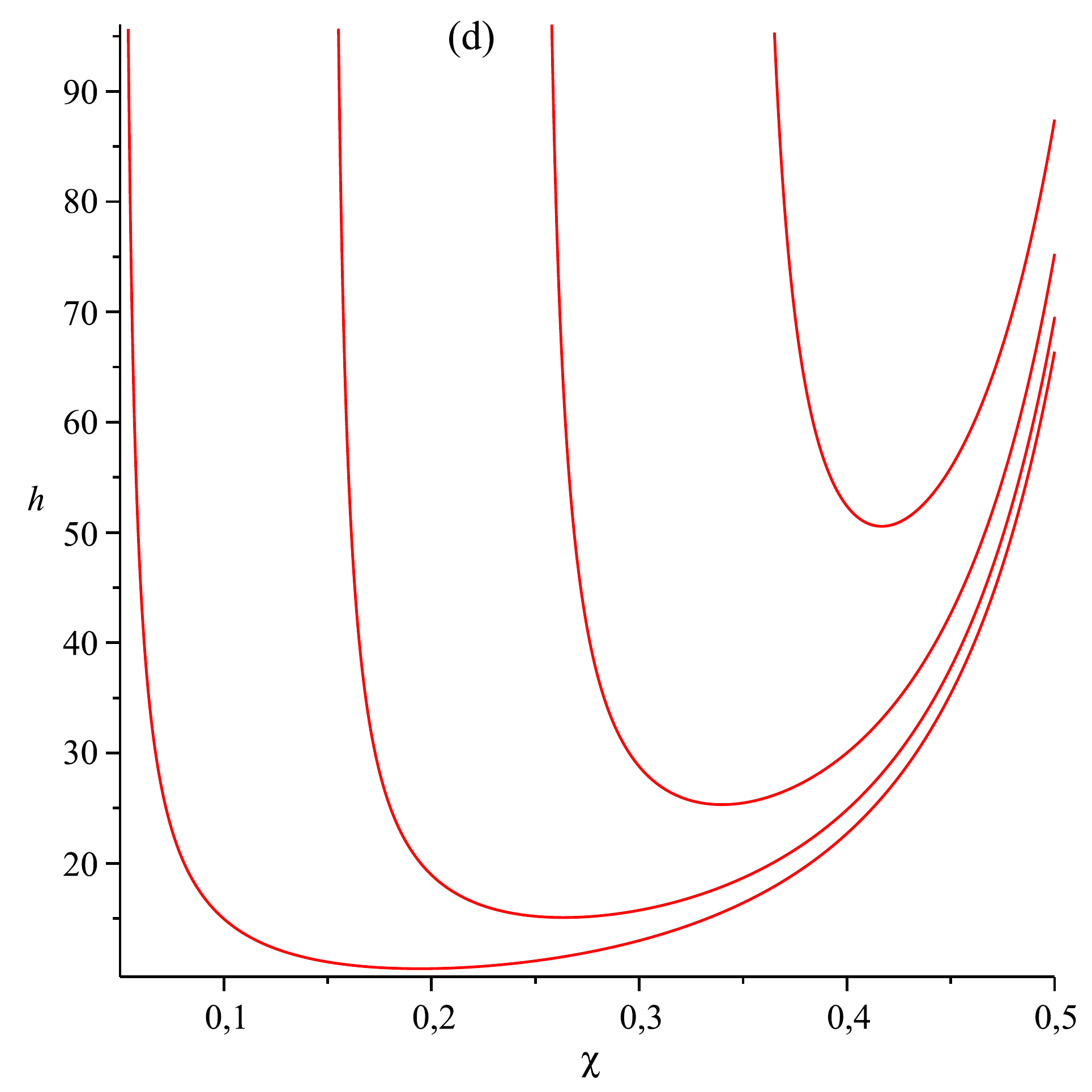}
\caption{\label{fig:nh} \small Metric components $\ln{N}$ and $h$ for the  classes of solutions 1 and 2. Panels (a) and (b) correspond to the class 1 and panels (c) and (d) to the class 2. In (a) and (b) each curve starts from the left at its corresponding $\chi_s^{(2)}$ singularity, passes through the throat $\hat{\chi}$ and continue towards $\chi\rightarrow +\infty$. $N$ and $h$ are completely regular in the whole domain $\chi\in(\chi_s^{(2)},+\infty)$ where these solutions are defined, which includes the throat. $h$ diverges as $\chi$ goes to $\chi_s^{(2)}$, whereas $N$ remains finite there. At $\chi\rightarrow +\infty$ $N$ and $h$ go to zero. In (c) and (d) each curve starts from the left at its corresponding $\chi_s^{(2)}$ singularity and continues towards the fixed $\chi_s^{(1)}$ singularity, which is located at the right end of all curves. $N$ and $h$ are regular in the full domain $\chi\in(\chi_s^{(2)},\chi_s^{(1)})$. $h$ diverges at both singularities $\chi_s^{(1)}$ and $\chi_s^{(2)}$ whereas $N$ diverges only at $\chi^{(1)}_s$. The values of $\beta$ and $\Lambda$ are the same of Fig.~\ref{fig:allsolutions}.}
 \end{center}
\end{figure}

\begin{enumerate}

\item {\bf Class 1} As we have discussed the domain of these solutions is $\chi\in (\chi_s^{(2)},+\infty)$. The numerical integration shows that both functions $N$ and $h$ are completely finite and nonzero on this domain. In particular they are finite and nonzero at the minimum $\hat{\chi}$, which in the plots is approximately at $\hat{\chi} = 1.47$. This shows that the minimum $\hat{\chi}$ is a regular point where the coordinate $r$ fails, since it repeats its values around $\hat{\chi}$. Thus, there is a regular throat at $\hat{\chi}$. Unlike $r$, The coordinate $\chi$ is useful to parameterize the space around the throat. Since $N$ and $h$ are regular and nonzero in the inner domain $(\chi_s^{(2)},+\infty)$, we conclude that there are no horizons or essential singularities in this domain (we recall the reader that the transformation from $r$ to $\chi$ determined by Eq.~(\ref{diffcoordtransf}) was purely on the radial direction, with no dependence in time involved). Thus, any geodesic passes from one side of the throat to the other one without reaching horizons or essential singularities at the inner points. There is a minimal-area two-sphere of $r=\mbox{constant}$ at the throat and the area of the spheres grows at the two sides of the throat. 

For $\chi$ higher than the throat the chart extends itself covering one side up to the spatial infinity $r=\infty$ $(\chi = +\infty)$. From the plots in Fig.~\ref{fig:nh} (a),(b) we see that $N^2$ and $h$ go simultaneously to zero at this asymptotic limit. We shall see that the Kretschmann scalar diverges at this limit, showing that this is an asymptotic essential singularity. The behavior of $N^2$ and $h$ at this asymptotic limit is completely equivalent to what happens in the infinite boundary of the inner side of the $\Lambda = 0$ solution \cite{Eling:2006df,Bellorin:2014qca}. 

In the other side of the throat, $\chi\in (\chi_s^{(2)},\hat{\chi})$, the solutions stop at the second-kind singularity. From Eq.~(\ref{hcomponent}) we see that $h(\chi)$ diverges at $\chi_s^{(2)}$ since $H = 0$ there, which is confirmed in the plots. We did not find any divergence or vanishing of $N$ at $\chi_s^{(2)}$ numerically. Since $N$ behaves as a scalar under pure spatial transformations, a feasible way to elucidate whether $\chi_s^{(2)}$ is a coordinate singularity is to go back to the original coordinate $r$, under which $f^{-1}$ is the corresponding radial component of the metric. We evaluate the function $f$ numerically using (\ref{sqrtf}). We find that $f$ is finite and nonzero at $\chi_s^{(2)}$, which we shall show explicitly in a procedure of joining solutions. The regularness of $N$ and $f$ at $\chi_s^{(2)}$ shows that $\chi_s^{(2)}$ is a coordinate singularity and $r$ is a good coordinate to cover this point (notice that the validity of the coordinates goes in opposite ways between $\hat{\chi}$ and $\chi_s^{(2)}$).

\item {\bf Class 2} The domain of these solutions is $\chi\in(\chi_s^{(2)},\chi_s^{(1)})$; $N$ and $h$ are finite and nonzero on this domain. Thus, there are no horizons or naked singularities in inner points. These solutions cover regularly an open subset of the space from the spatial infinity $r = \infty$ $(\chi = \chi_s^{(1)})$ down to a finite point $r = r_s^{(2)}$ $(\chi = \chi_s^{(2)})$. Both $N(\chi)$ and $h(\chi)$ diverge at the right of the plots, where the first-kind singularity $\chi_s^{(1)}$ is reached. This leads us to ask ourselves whether this asymptotic divergence follows the deformed AdS asymptotia we discussed previously.

At the left end, where the solutions stop at $\chi_s^{(2)}$, $h$ diverges again whereas $N$ remains finite and nonzero. Thus, we have that the behavior of this solution at the second-kind singularity is very similar to the previous class. We find that $f(\chi_s^{(2)})$ is finite and nonzero, hence for these solutions $\chi_s^{(2)}$ is also a coordinate singularity.

\item{\bf Class 3} The domain of these solutions is $\chi\in(-\infty,\chi_s^{(1)})$; $N$ and $h$ are finite and nonzero over this domain. These solutions cover the full space $r \in (0,\infty)$ in a regular fashion, there are no horizons or essential singularities in inner points, and there are no critical points. At the limit $r\rightarrow 0$ $N$ diverges and $h$ goes to zero. The divergence of $N$ leads us to think that this is an essential singularity, which we are going to confirm with the curvature. At the limit $r\rightarrow \infty$ we have that $N^2 , h|_{r=\infty} = \infty$, thus we shall look for deformed AdS asymptotia also in this solution. 

\end{enumerate}


\subsubsection{The joining of two solutions}
The fact that solutions of classes 1 and 2 end regularly below and above of some finite $r$ suggests that both solutions can be joined to form a single wormhole geometry composed of two sides joined by a throat (the throat class 1 has), with its two sides extending themselves to spatial infinity, as the wormhole solution of the $\Lambda = 0$ case \cite{Eling:2006df,Bellorin:2014qca}.

\begin{figure}[t]
 \begin{center}
 \includegraphics[scale=0.4]{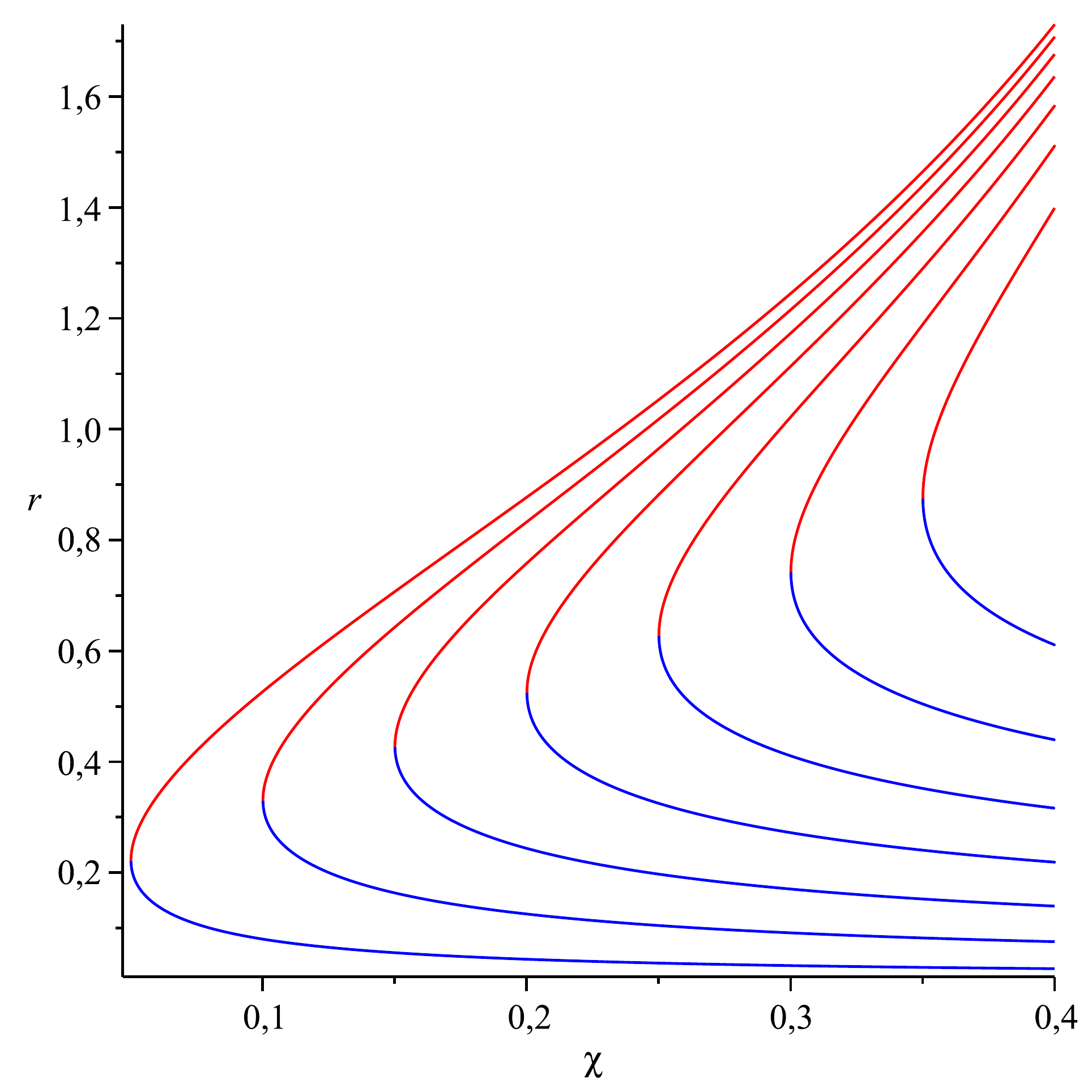}
 \caption{\label{fig:joinning} \small Curves of classes 1 (blue) and 2 (red) that tend to the same second-kind singularity $\chi_s^{(2)}$, plotted near the singularity. For each joined curve the $\chi_s^{(2)}$ singularity corresponds to the joining point of its red and blue parts.}
 \end{center}
\end{figure}

To clearly show the joining, we generate solutions $r(\chi)$ of classes 1 and 2 that tend to the same $\chi_s^{(2)}$ singularity by giving initial data in the form
\begin{equation}
 r(\chi_s^{(2)} + \delta) = r_s^{(2)} \pm \epsilon \,.
\label{nearsingularity}
\end{equation}
For the case $-\epsilon$ we obtain curves with the throat $\hat{\chi}$ (class 1) and for $+\epsilon$ we obtain the intermediate curves that have both kind of singularities (class 2). For $\delta$ and $\epsilon$ sufficiently small we get that both curves tend to join themselves at $\chi_s^{(2)}$; such that the image of their union, including the value $r^{(2)}_s$, constitutes an extended, continuous, and doubly copied range of $r$. In Fig.~\ref{fig:joinning} we show such continuous joinings of the functions $r(\chi)$. We generate several curves by varying $\chi_s^{(2)}$ in the range $(0,\chi_s^{(1)})$. We see that in all cases the coordinate $r$ is regular and continuous at the joining point.

\begin{figure}[!ht]
 \begin{center}
 \includegraphics[scale=0.34]{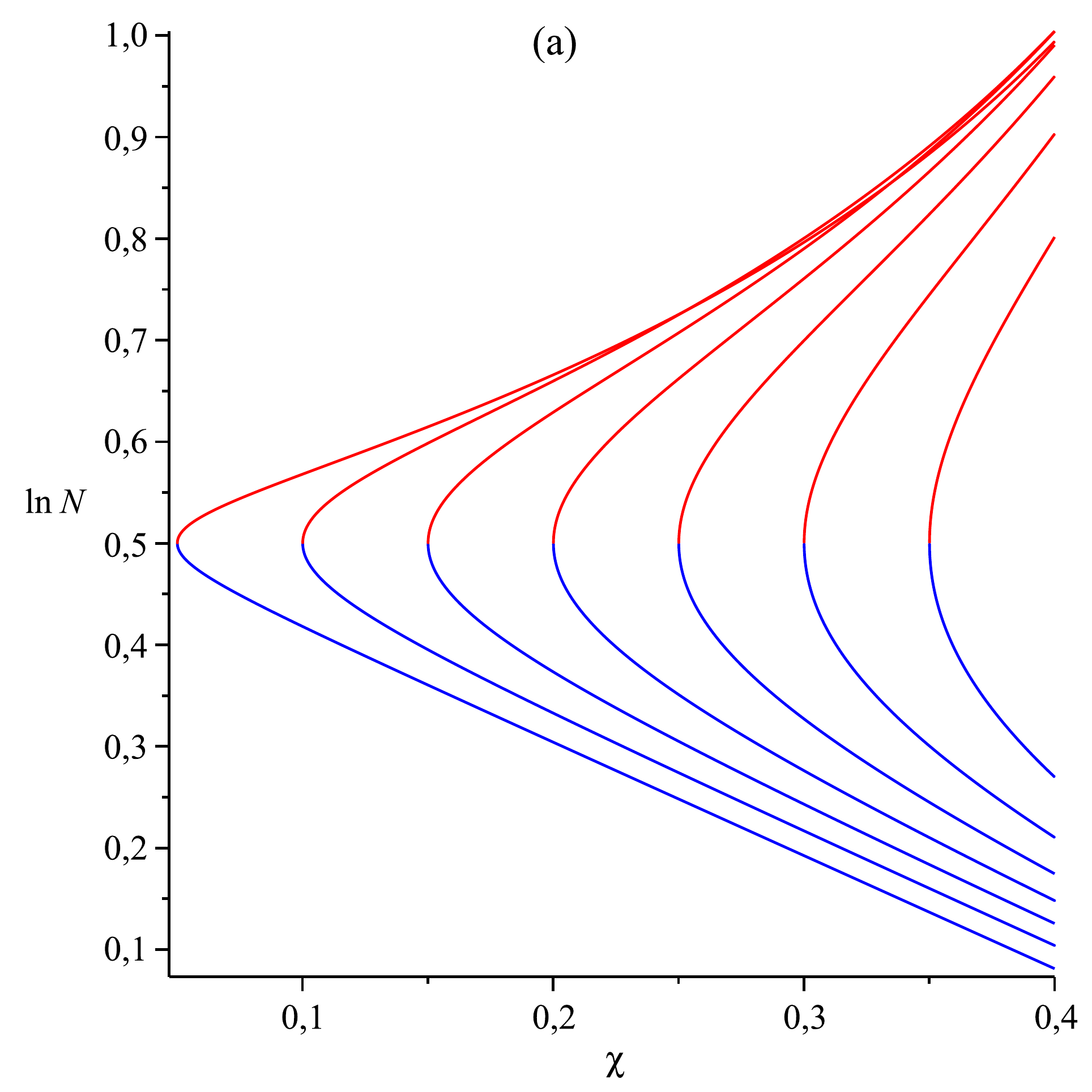}
 \includegraphics[scale=0.34]{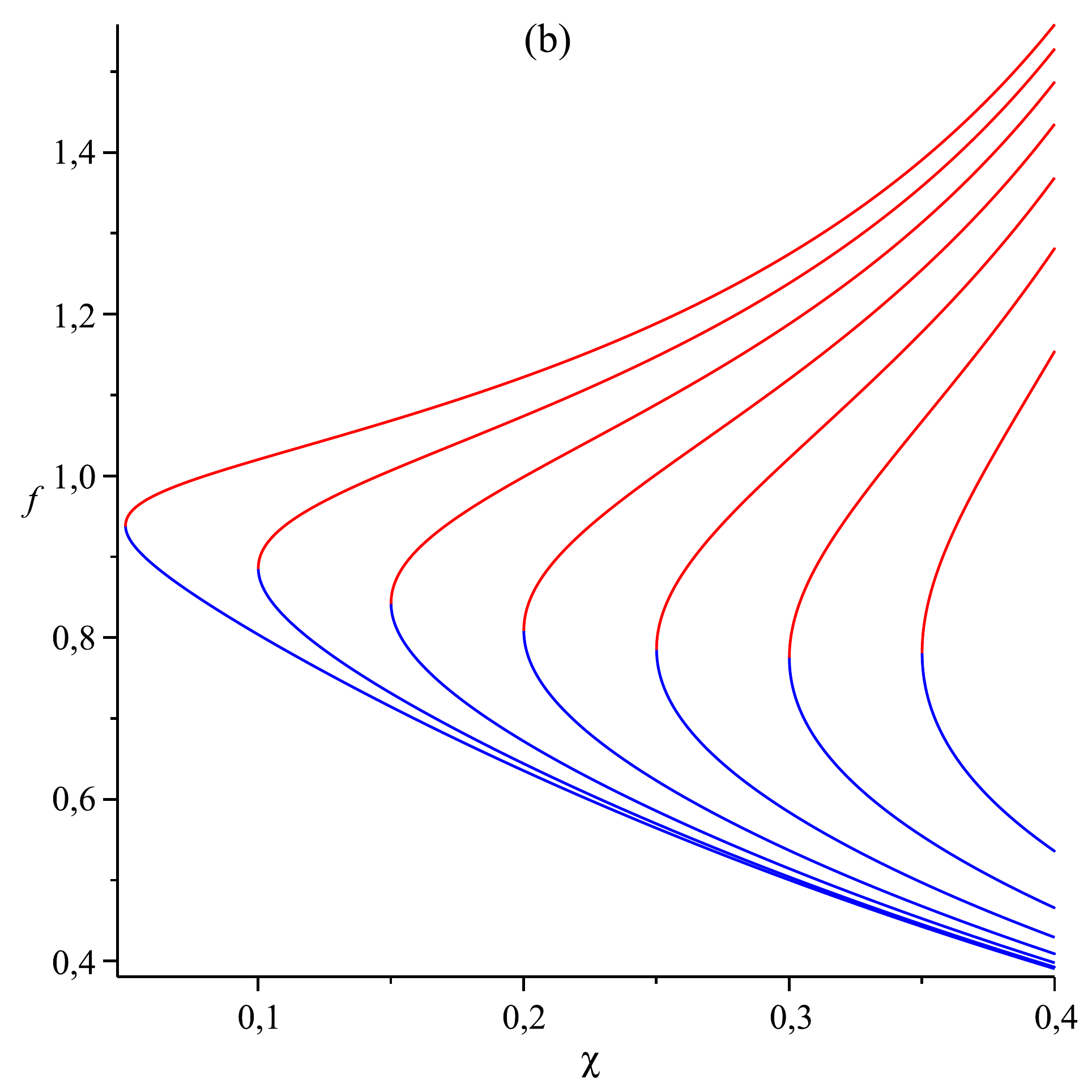}
 \includegraphics[scale=0.34]{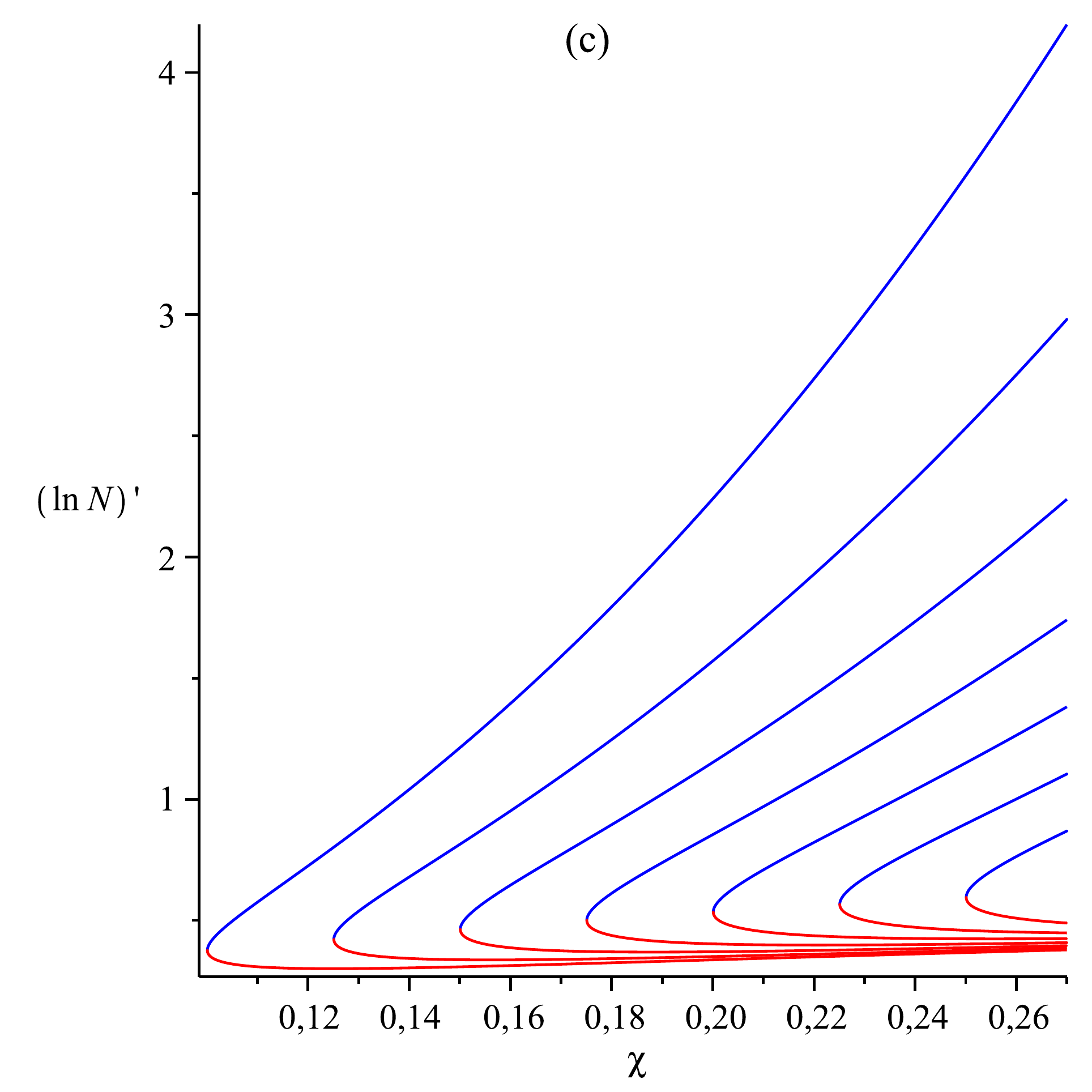}
 \includegraphics[scale=0.34]{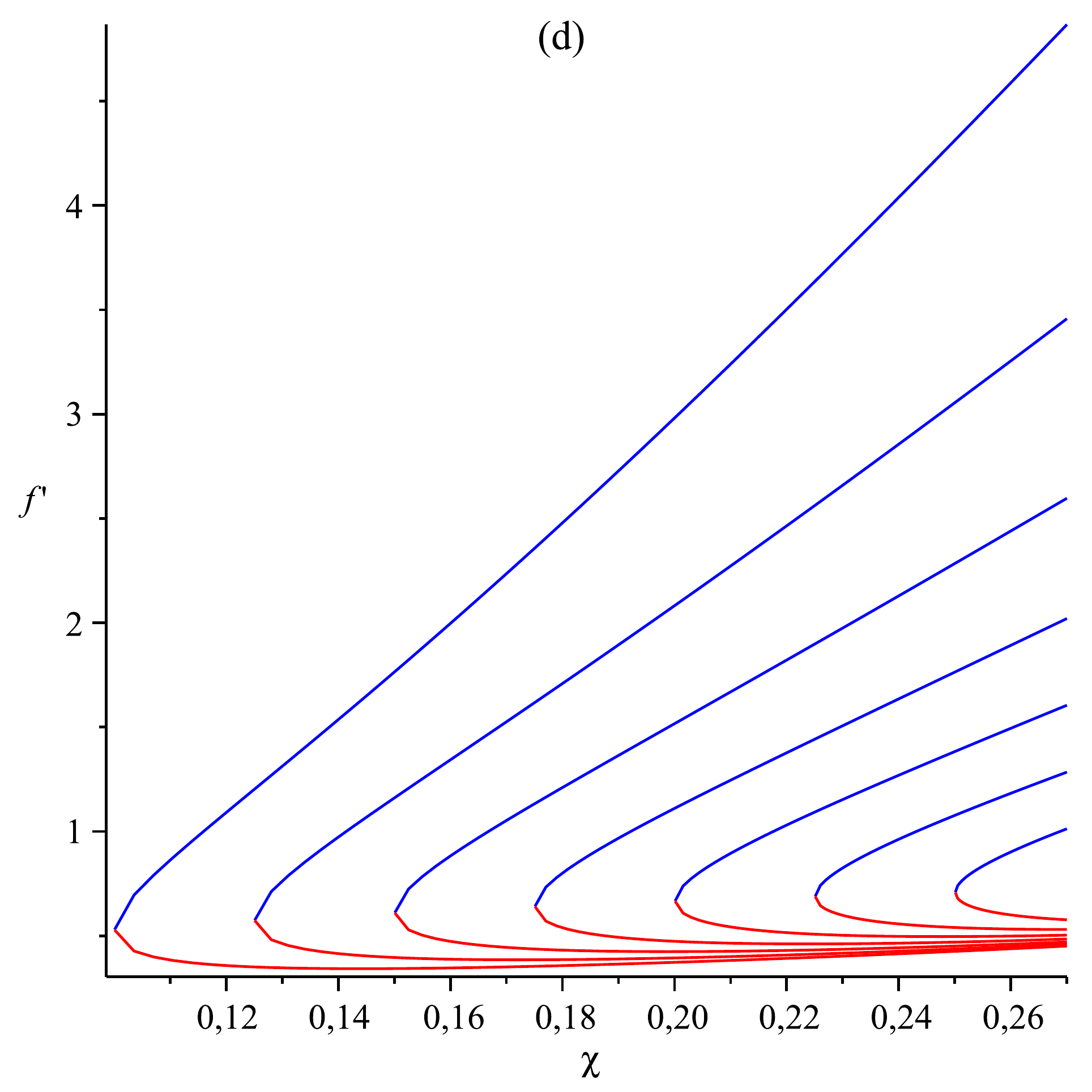}
 \caption{\label{fig:joinningnf} \small Functions (a) $\ln{N}$, (b) $f$, (c) $d\ln{N}/dr$ and (d) $df/dr$  for solutions of classes 1 (blue) and 2 (red) near the second-kind singularity. All of these functions remain continuous as functions of $r$.}
 \end{center}
\end{figure}

We may also see continuity and smoothness in the metric components. The derivative $d\ln{N}/dr$ is taken from Eq.~(\ref{nprime}) and $df/dr$ follows by taking a derivative on Eq.~(\ref{sqrtf}) with respect to $r$ and combining with Eq.~(\ref{diffcoordtransf}). This yields
\begin{equation}
 \frac{df}{dr} =
    2 \sinh{\chi} (\beta^{-1} \cosh{\chi} - \sinh{\chi})
     \frac{H}{r} 
  - 2 \Lambda \left( \cosh{\chi} - \beta^{-1} \sinh{\chi} \right)^2 r \,.
\label{dfdr}
\end{equation}
Thus, Eqs.~(\ref{nprime}) and (\ref{dfdr}) give $d\ln{N}/dr$ and $df/dr$ as functions of $\chi$. In Fig.~\ref{fig:joinningnf} we plot the numerical solutions for $N(\chi)$, $f(\chi)$ and their derivatives using again (\ref{nearsingularity}) combined with $N(\chi_s^{(2)} + \delta) = N_0 \pm \epsilon$, where $N_0$ is a fixed value (this is equivalent to adjust the integration constants of both solutions). We may see that $N$, $f$ and their derivatives are finite, nonzero and completely continuous at the joining point. Thus we conclude that solutions of the classes 1 and 2 form a single solution that is regular at the joining point $r_s^{(2)}$. We remark again that the passage from $\chi$ to $r$ is a pure spatial transformation involving only the radial direction. Thus, there is no horizon at the joining point. We have already discussed that the two separated solutions of classes 1 a 2 have no horizons or essential singularities at finite points.

The theory has two static spherically symmetric solutions with vanishing shift function: (i) a wormhole solution whose geometry consists of two sides, nonsymmetric between them, joined by a throat and with the two sides extending themselves from the throat to the spatial infinity. This solution is completely regular in its inner points, there are no horizons or essential singularities at finite locations. (ii) A solution that covers a single copy of the space from $r=0$ to $r=\infty$ and that is regular in $r\in(0,\infty)$. There remains to analyze the asymptotic limits on the two sides of the wormhole and the behavior of the last solution at $r=0$ and $r=\infty$, as well as the inspection of the curvature at all of these points.


\subsubsection{Numerical asymptotia and curvature}
We contrast with the Schwarzschild-anti-de Sitter space and check the asymptotia described in the preliminary analysis. The Schwarzschild-anti-de Sitter space arises in our approach by putting $\beta = 1$ ($\alpha = 0$) in Eqs.~(\ref{diffcoordtransf}), (\ref{dNdchi}) and (\ref{sqrtf}). All the analysis of the singularities of Eq.~(\ref{diffcoordtransf}) still holds with $\beta = 1$. Thus, there is a singularity of the first kind at $\tanh{\chi_s^{(1)}} = 1/2$ and singularities of the second kind in $(0,\chi_s^{(1)})$. Again, two solutions ending at the same $\chi_s^{(2)}$ can be smoothly joined to form  single solution. When $\beta = 1$ the critical point $\hat{\chi}$ moves to infinity, so there is no throat at all, as expected. Instead, for $\chi \rightarrow +\infty$ Eq.~(\ref{diffcoordtransf}) implies $dr/d\chi \rightarrow 0^{-}$. This implies that the function $r(\chi)$ has a lower asymptote at $\chi\rightarrow +\infty$. This asymptote is the location of the horizon, hence this solution in the $\chi$ coordinate covers the exterior region of the Schwarzschild-anti-de Sitter space. This fact is corroborated in Eq.~(\ref{rinfinity}), since $\tilde{n} = 0$ when $\beta = 1$, meaning that $r=\infty$ is reached only at $\chi = \chi^{(1)}_s$ in this case. The class 3 solution, which extends itself to $\chi \rightarrow -\infty$, has the essential singularity (the Schwarzschild-anti-de Sitter space with negative mass). These results are completely analogous to the way we recovered the Schwarzschild solution in the $\Lambda = 0$ case \cite{Bellorin:2014qca}.

\begin{figure}[!t]
 \begin{center}
 \includegraphics[scale=0.34]{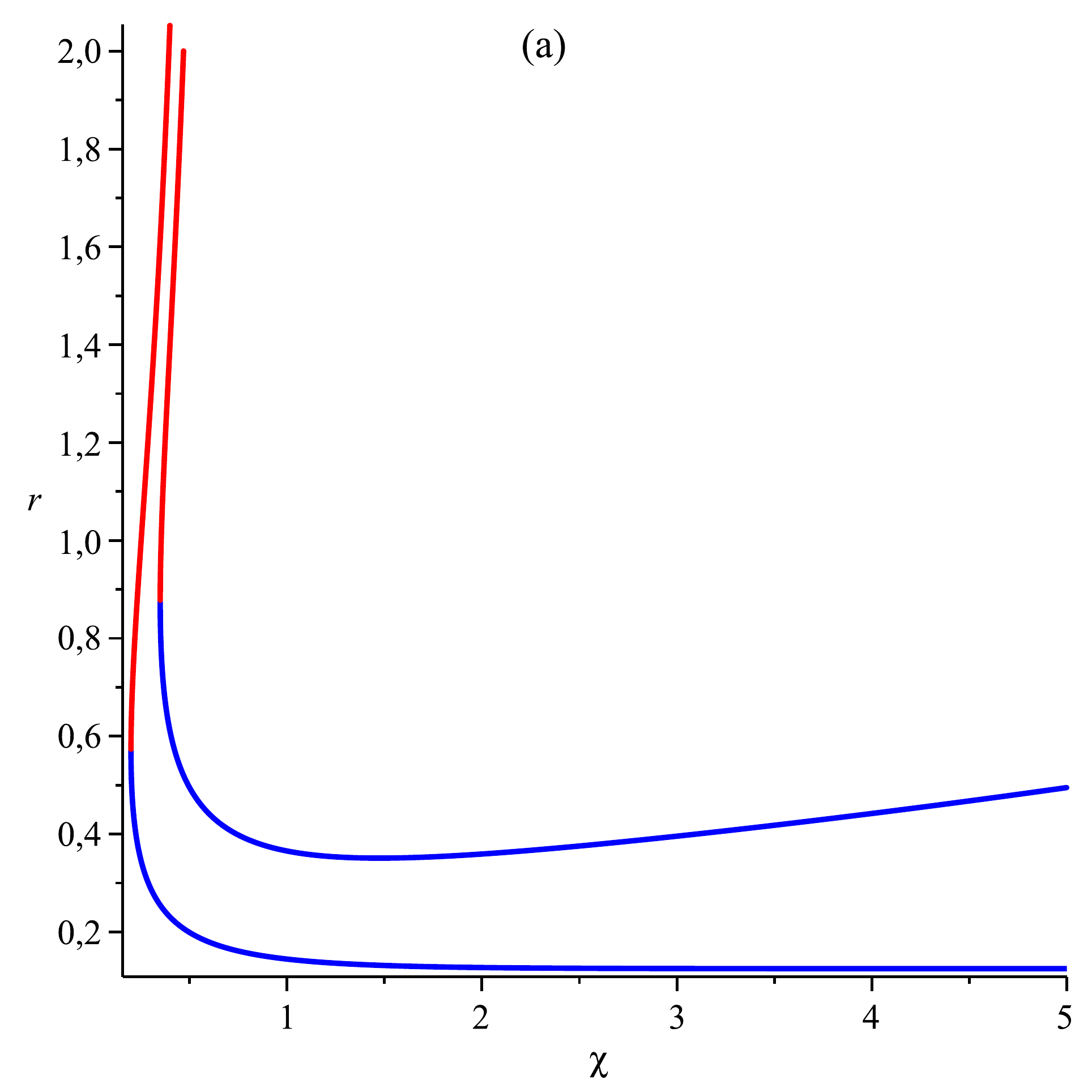}\\
 \includegraphics[scale=0.34]{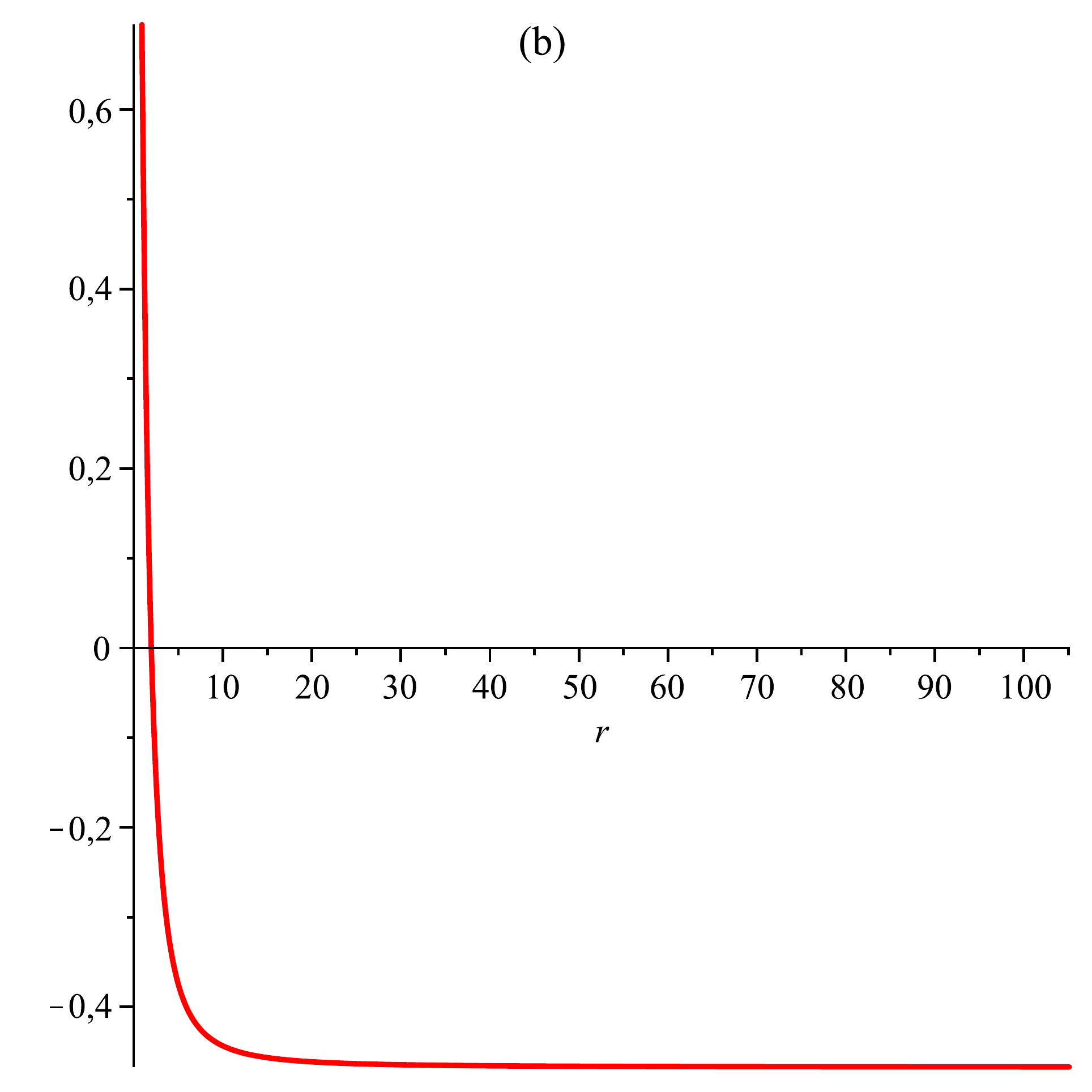}
 \includegraphics[scale=0.34]{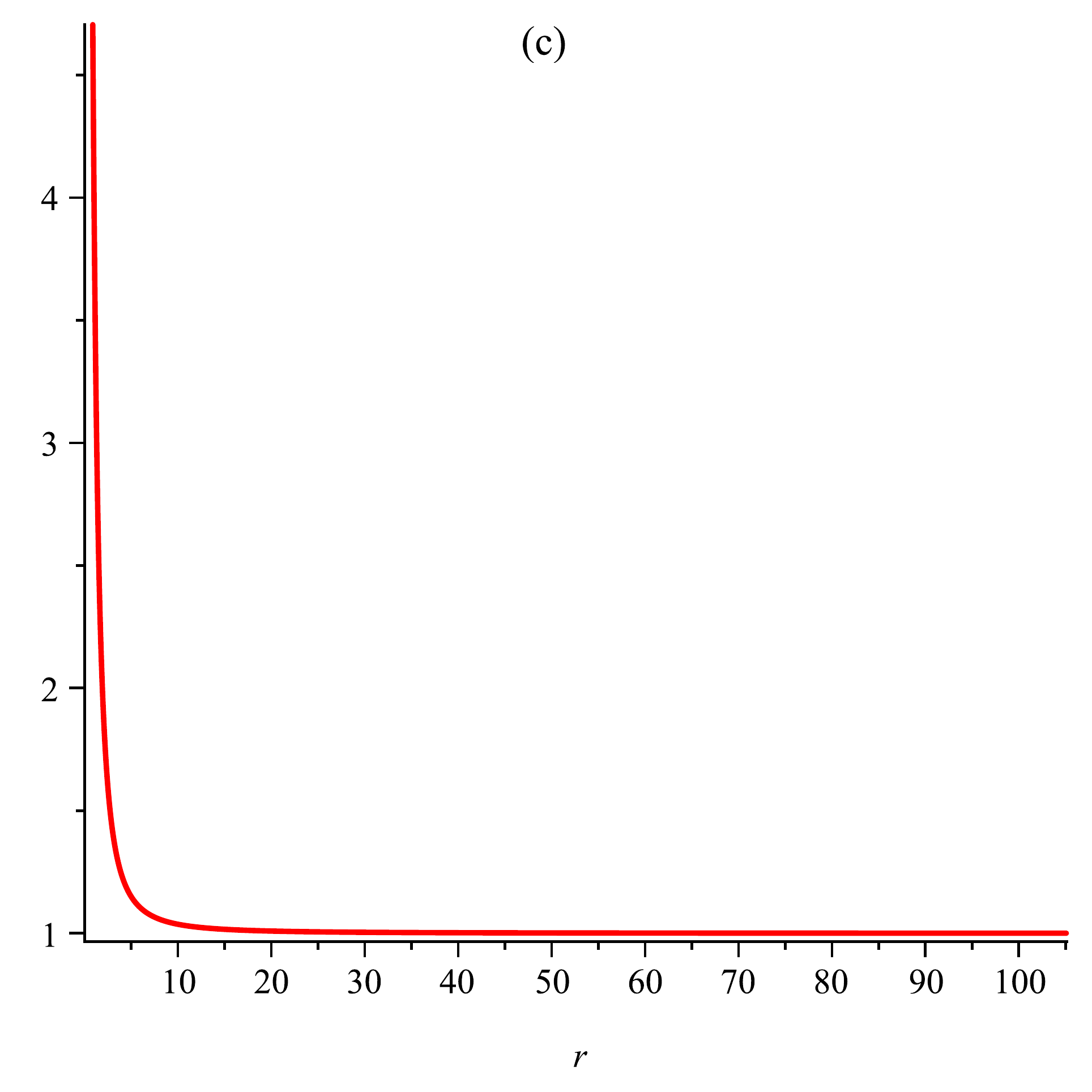}
 \caption{\label{fig:ads} \small (a) Function $r(\chi)$ with $\alpha = 0$ (lower curve) and $\alpha \neq 0$ (upper curve) in the range $\chi\in(\chi_s^{(2)},+\infty)$. The $\alpha \neq 0$ solution has a minimum, hence it has the throat, whereas the $\alpha = 0$ solution has a lower asymptote; it is the exterior of the Schwarzschild-anti-de Sitter black hole. In (b) and (c) we compare $N$ and $f$ of one side of the joined solution with the deformed AdS asymptotia. The curve in (b) is $\ln{(N(r)/N_0(r))}$. It goes strongly to a constant value as $r$ increases. The curve in (c) is $f(r)/f_0(r)$. It goes strongly to one.}
 \end{center}
\end{figure}

In Fig.~\ref{fig:ads} (a) we show the functions $r(\chi)$ for a wormhole solution and for the (positive mass) Schwarzschild-anti-de Sitter space. Each solution has been joined at its respective $\chi_s^{(2)}$ singularity. In general these solutions are very different since the wormhole covers two copies of a subset of the space, whereas the exterior black hole covers only one. Looking at the divergences we have found so far, it is clear that the side of the joined wormhole solution that has some similarity with the exterior black hole is the one containing the intermediate solution (class 2). In Figs.~\ref{fig:ads} (b), (c) we compare the asymptotic behavior of the functions $N(r)$ and $f(r)$ on this side for a particular class 2 solution with respect to the asymptotia found from (\ref{nfdiverging}) to (\ref{ac}); that is, we compare with $N_0 = r^{(1-\alpha)^{-1}}$ and $f_0 = C r^2$, where $C$ is given in (\ref{ac}). In the plot (c) we see that the ratio $f/f_0$ goes very strongly to one. In (b) the logarithm of the ratio, $\ln{N/N_0}$, also goes very strongly to a constant, which is the expected behavior. The value of this constant does not need to be zero since it is the coefficient of the dominant mode of $N$, which can be regarded as an integration constant (this freedom is not present in $f$). Therefore, the plots in Figs.~(\ref{fig:ads}) confirm the asymptotic behavior $N^2 \sim r^{2(1-\alpha)^{-1}}$ and $f\sim Cr^{2}$ in one of the sides of the wormhole solution. For small $\alpha$, which is the case we have considered throughout all the analysis, we interpret this asymptotia as a deformed AdS space. In the other side, as we have already pointed out, $N^2,h\rightarrow 0$ asymptotically. For the class 3 solution we have also checked that there is deformed AdS asymptotia as $r\rightarrow\infty$.

\begin{figure}[t]
 \begin{center}
 \includegraphics[scale=0.4]{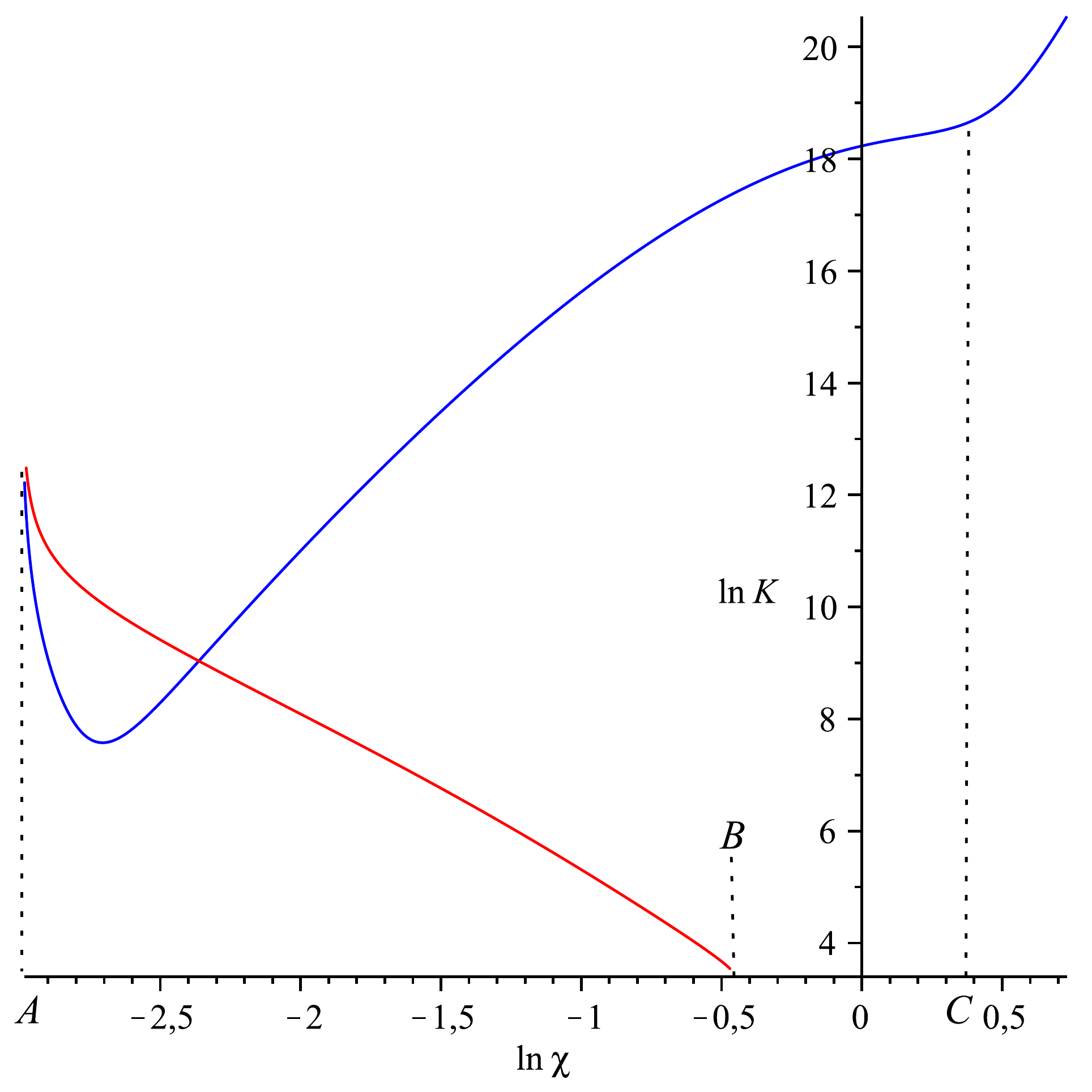}
 \caption{\label{fig:kretschamnn} \small Kretschamnn scalar for the wormhole solution. The indicated points are $A = \ln{\chi_s^{(2)}}$ (the joining point), $B = \ln{\chi_s^{(1)}}$ and $C = \ln{\hat{\chi}}$.  We see that $\mathcal{K}$ is finite and continuous at the joining point. The asymptotic limit of the side of the solution that is asymptotically deformed AdS can be seen in the red curve, for which $\mathcal{K}$ tends to a finite value as $r\rightarrow\infty$ ($\chi\rightarrow \chi_s^{(1)}$). The asymptotic limit in the other side can be seen in the blue curve, for which $\mathcal{K}$ diverges as $r\rightarrow\infty$ ($\chi\rightarrow +\infty$). Thus, there is curvature singularity at this asymptotic limit. In this solution $\mathcal{K}$ is regular over the whole solution, except at the essential asymptotic singularity at $\chi\rightarrow +\infty$.} 
 \end{center}
\end{figure}

The regularness of the curvature of the wormhole solution at the joining point $\chi_s^{(2)}$ as well as the potential presence of essential singularities in both solutions can be inspected in the Kretschmann scalar $\mathcal{K} = R_{\mu\nu\alpha\beta} R^{\mu\nu\alpha\beta}$. In the $\chi$ coordinate it takes the form
\begin{equation}
\begin{array}{rcl}
 \mathcal{K} &=& 
 {\displaystyle \frac{1}{N^2 h^2} \left( 2\frac{d^2 N}{d\chi^2} 
   - \frac{1}{h}  \frac{dh}{d\chi} \frac{dN}{d\chi} \right)^2
  + \frac{8}{r^2 N^2 h^2}\left( \frac{dN}{d\chi} \frac{dh}{d\chi} \right)^2 }
\\[1ex] &  &
 {\displaystyle  + \frac{2}{r^2 h^2} \left( 2 \frac{d^2 r}{d\chi^2}
       - \frac{1}{h} \frac{dh}{d\chi} \frac{dr}{d\chi} \right)^2
  + \frac{4}{r^4 h^2} \left( \left( \frac{dr}{d\chi} \right)^2 - h \right)^2 } \,,
\end{array}
\end{equation}
where $r$, $N$ and $h$ are understood as the functions of $\chi$ that determine the metric components in (\ref{metricchi}). In Fig.~\ref{fig:kretschamnn} we show the plot for the joined wormhole solution (classes 1 and 2). 

The first feature we remark is that the scalar $\mathcal{K}$ is finite and continuous at the joining point. This confirms that this point is just a failure of the $\chi$ coordinate. Second, the scalar $\mathcal{K}$ is regular everywhere except at $\chi\rightarrow +\infty$. At $\chi\rightarrow +\infty$ the spatial infinity on the inner side is reached. We observe that $\mathcal{K}$ diverges at this limit, thus we conclude that there is an essential singularity at this asymptotic limit. At the asymptotic limit of the other side $\mathcal{K}$ tends to a finite value, which is the behavior expected for the deformed AdS asymptotia. In the solution of class 3 we checked that $\mathcal{K}$ diverges at $r=0$ ($\chi = -\infty$), thus there is a naked essential singularity at the origin in this solution, whereas at the $r\rightarrow \infty$ limit ($\chi \rightarrow \chi_s^{(1)}$) $\mathcal{K}$ tends to a finite value which corresponds to the deformed AdS asymptotia, similarly to the exterior side of the previous solution.

\section{Conclusions}
We have focused the problem of finding the static spherically symmetric solutions of the Ho\v{r}ava theory \cite{Horava:2009uw,Blas:2009qj} with a negative cosmological constant. We have taken the lowest order effective action of the complete nonprojectable theory and we have switched off the shift function in the spacetime metric (with only the symmetry of foliation-preserving diffeomorphisms available the spherical symmetry does not imply staticity and these are not enough to have a vanishing shift function). We have assumed a small $\alpha$, which is the coupling constant of the $(\partial\ln{N})^2$ term. This term is the unique deviation from GR in the effective action \cite{Bellorin:2010je}, thus we are interested in a small $\alpha$.  We have carried out a handling of the field equations similar to the one of Ref.~\cite{Bellorin:2014qca}, where the same problem was focused for the case of vanishing cosmological constant. As it happened in that paper, here we have found a minimum for the physical radius $r$ when it is regarded as a function of a new coordinate that runs at both sides of the minimum. This corresponds to a wormhole-like geometry of two sides joined by a throat, the minimum being the throat. The two sides are nonsymmetric between them and each one extends itself from the throat to the spatial infinity. The wormhole solution is everywhere regular and without horizons, except for an asymptotic essential singularity in one of the sides

Curiously (and different to the $\Lambda = 0$ case), the new coordinate, $\chi$, fails at a certain finite point of one side. We have identified this as a coordinate singularity. Thus, nothing special occurs at that point but a failure of the $\chi$ coordinate. Instead, the physical radius $r$ is regular and valid there. We supported this conclusion by examining the metric components numerically and the square Riemann tensor (Kretschmann scalar).

In the wormhole solution of the $\Lambda = 0$ case \cite{Eling:2006df,Bellorin:2014qca} the asymptotia at the two sides is different. One side is asymptotically flat whereas the other one is asymptotically singular. The singular side can be regarded as a kind of inner space. In the case we studied here we found an analogous nonsymmetric asymptotic behavior, but with some differences. Asymptotically, one wish to compare the exterior side with AdS, which has $N^2,f\sim r^2$ asymptotically. However, we found analytically that the divergence of this side is of the form $N^2 \sim r^{2(1-\alpha)^{-1}}$, $f\sim r^2$. This was confirmed by the numerical analysis. Since $\alpha$ is considered small, we interpret this result as a deformed-AdS asymptotia, characterized by a growing of $N^2$ faster than AdS. The other side, which we regard as an internal space, exhibits an asymptotic essential singularity, as in the $\Lambda = 0$ case.

Interestingly, the asymptotic behavior of the regular side is very similar to the Lifshitz scaling solution, which has been used in the context of holographic duals of nonrelativistic configurations. In the context of the Ho\v{r}ava theory, in Ref.~\cite{Griffin:2012qx} it was shown that the Lifshitz scaling spacetime is a vacuum solution of the complete nonprojectable theory with cosmological constant. To make a comparison with our solution, the ``bulk'' coordinate of the Lifshitz spacetime ($r$ in the notation of \cite{Griffin:2012qx}) can be equated to the radial direction we have used here. Although the spatial submanifolds of $r = \mbox{constant}$ are different in both cases, the asymptotic $tt$ and $rr$ components of our solutions in the nonsingular side are exactly equal to the ones of the Lifshitz spacetime. In particular, the exponent of our $-N^2$ function, which motivates the name of deformed AdS, is equal to the exponent of the analogous component in the Lifshitz spacetime, which leads to the anisotropic scaling. Thus, asymptotically our solutions acquire a kind of Lifshitz scaling.

Whole wormhole solutions with negative cosmological constant have been studied in GR and its extensions in many contexts. For example, they have been found in combination with scalar fields \cite{Anabalon:2012tu}, their quantization has been carried out \cite{Barcelo:1995gz} and they has been used in the AdS/CFT correspondence \cite{Maldacena:2004rf}. We hope that the solutions of the Ho\v{r}ava theory can find similar applications. 

The side with deformed AdS asymptotia can be used as exterior solution of stars, as was done in Ref.~\cite{Eling:2006df} for the analogous solution in the $\Lambda = 0$ case. This consists of taking the deformed-AdS side of the wormhole and truncating it at some finite radius greater than the location of the throat. Since the distribution of matter is located from the origin to the radius of joining, the vacuum solution is not valid inside the joining surface. Thus, such solutions have neither throat nor internal side at all. As exterior solution, it is interesting to have an asymptotic behavior slightly different to the one of AdS. 

Returning to the whole wormhole solution, an important feature it has is that it is a \emph{vacuum} solution. In GR it is well known that exotic matter is  needed to maintain a traversable wormhole. But in the solution we have at hand this is not the case, its existence does not require the coupling to matter. Geodesics pass from one side to the other one through the throat without seeing any matter, neither exotic nor ordinary. Thus, no exotic matter is needed to maintain this solution. The same feature holds for the wormhole solution of the $\Lambda = 0$ case. This is conceptually similar to some traversable wormholes that have been found in the braneworld scenario \cite{Bronnikov:2002rn,Lobo:2007qi}, where the field equations allow the space to maintain its throat by itself, with no need of matter violating energy conditions (in the braneworld it is the effect of the bulk geometry that maintains the throat on the brane). 

We have also found solutions with naked singularities, like in the $\Lambda = 0$ case. In particular, only such solutions arise in the case of negative $\alpha$. These solutions cover a single copy of the $r\in(0,\infty)$ space and also describe deformed AdS asymptotia.

About the name that can be given to our two-side solution, we comment that the notion of a wormhole geometry is a global one, it is not sufficient to find a throat, which is a local feature, to conclude that some space can be classified as a wormhole. In particular there should no be physical barriers avoiding to pass from one side two the other one in the two directions through the throat. An study on these issues for some solutions was carried out in Refs.~\cite{Bronnikov:2006pt,Bronnikov:2010tt}. It was found that some solutions of the Brans-Dicke theory and the $f(R)$ theory proposed as wormholes are actually global wormholes only under conditions for which the theory propagates ghosts. In some instances configurations named as wormholes are considered with diverse asymptotia, see \cite{Hochberg:1996ee,Hochberg:1997wp}. The numerical analysis we have performed here constitutes a global study of the configurations. We have found no barriers that would avoid the two-way movement of physical particles between the internal points of the two sides of the solutions (excluding the asymptotic singularity at one side). We also comment that the complete nonprojectable Ho\v{r}ava theory at $\lambda = 1/3$ has no ghosts (it has no extra degrees at all) \cite{Bellorin:2013zbp} and for $\lambda \neq 1/3$ one may require $\lambda > 1$ in the linearized theory to safe the extra degree from becoming a ghost \cite{Blas:2009qj}. The solutions we have found here are valid for any $\lambda$.

Finally, the stability of our solutions under small perturbations is an important aspect that deserves a detailed analysis. Such an investigation is beyond the scope of this paper.


\section*{Acknowledgments}
A. R. and A. S. are partially supported by Project Fondecyt No.
1121103, Chile.

\appendix
\section*{Appendix: Negative $\alpha$}
When $\alpha$ is small and negative, $\alpha \sim 0^-$, we have that $\beta \sim 1^+$. In this case all the analysis remains identical up to the Eq. (\ref{rlambdacero}). The singularities of Eq.~(\ref{diffcoordtransf}) are very similar. There are first-kind singularities at $\tanh{\chi_s^{(1)}} = 1/2\beta$ and second-kind singularities in $(0,\chi_s^{(1)})$. The interpretation of both singularities is the same: they are coordinate singularities. Thus, the two solutions ending at the same $\chi^{(2)}_s$ can be joined to form a single solution. Therefore, there are two solutions: one formed by the joining of two solutions in $(\chi^{(2)}_s,+\infty)$ and the other one in $(-\infty,\chi^{(1)}_s)$.

The sector $\chi \in (-\infty,\chi^{(1)}_s)$ is very similar to the $\alpha > 0$ case. These solutions have neither second-kind singularities nor critical points. They have an essential singularity at the origin and cover $r\in (0,\infty)$. We confirmed this by evaluating numerical solutions.

The sector $(\chi^{(2)}_s,+\infty)$ is very different to the $\alpha>0$ case. The derivative $dr(\chi)/d\chi$ at $\chi\rightarrow +\infty$ becomes negative. Thus, $r$ becomes monotonically decreasing at $\chi\rightarrow +\infty$, such that the spatial infinity is not reached at $\chi\rightarrow +\infty$, but instead the solution goes to the origin. This is confirmed by Eq.~(\ref{rinfinity}), since now $\tilde{n} < 0$. Moreover, there is no possibility of critical point here (Eq.~(\ref{throat}) has no solution). This implies that this part of the solution is monotonically decreasing in the full domain $(\chi^{(2)}_s,+\infty)$. Hence, its union with the other part covers the full range $r\in (0,\infty)$. It has a naked essential singularity at the origin. We confirmed this result with numerical integration.

Therefore, in the $\alpha \sim 0^-$ case all the static spherically symmetric solutions with vanishing shift function have naked singularities at the origin and cover a single copy of the full space $r\in(0,\infty)$. They acquire deformed AdS asymptotia as $r\rightarrow\infty$.


\end{document}